\documentclass[journal=./achemso/jpc,manuscript=article,layout=onecolumn]{achemso}
\usepackage{geometry}
\usepackage{amsmath}
\usepackage{amsfonts,amstext,amsgen,amsbsy,amsopn,amssymb}
\usepackage[braket,qm]{qcircuit}
\usepackage{graphicx}
\usepackage{dcolumn}
\usepackage{bm}

\usepackage{physics}
\usepackage{makecell}
\usepackage{booktabs}
\usepackage{caption}
\usepackage{subcaption}
\usepackage{multirow}
\usepackage[dvipsnames, table]{xcolor}

\definecolor{diagramColor}{RGB}{0,0,120}

\usepackage{tikz}
\usetikzlibrary{trees}
\tikzset{
 	treenode/.style = {shape=rectangle, draw, diagramColor, ultra thick,
        align = center,
        top color = white,
        bottom color = white},
        root/.style = {treenode, font=\normalsize},
        env/.style = {treenode, font=\normalsize},
        envtwo/.style = {treenode, font=\normalsize},
        mode/.style = {edge from parent path={(\tikzparentnode.east) -- (\tikzchildnode.west)}}
	}
\usepackage{geometry}
\usepackage{enumitem}
\usepackage{mathtools}
\usepackage[colorlinks=true,linkcolor=blue,citecolor=blue,urlcolor=black]{hyperref}

\usepackage{array}
\newcolumntype{?}{!{\vrule width 1.5pt}}

\newcommand{\halfcheckmark}[0]{\checkmark\raisebox{0.23em}{\kern-0.68em\large$\times$}}

\SectionNumbersOn

\title{Holographic Gaussian Boson Sampling with Matrix Product States on 3D cQED Processors}

\author{Ningyi Lyu}
\affiliation{Department of Chemistry, Yale University, New Haven, CT 06520, U.S.A.}

\author{Paul Bergold}
\affiliation{Department of Mathematics, University of Surrey, Guildford, U.K.}

\author{Micheline Soley}
\affiliation{Department of Chemistry, Yale University, New Haven, CT 06520, U.S.A.}
\alsoaffiliation{Yale Quantum Institute, Yale University, New Haven, CT 06511, U.S.A.} 
\alsoaffiliation{Department of Chemistry, University of Wisconsin-Madison, 1101 University Ave., Madison, WI 53706, U.S.A.}

\author{Chen Wang}
\affiliation{Department of Physics, University of Massachusetts-Amherst, Amherst, MA 01003, U.S.A.} 

\author{Victor S. Batista}
\affiliation{Department of Chemistry, Yale University, New Haven, CT 06520, U.S.A.}
\alsoaffiliation{Yale Quantum Institute, Yale University, New Haven, CT 06511, U.S.A.} 
\email{victor.batista@yale.edu}

\date{\today}
\makeatother

\begin{document}
\section{Abstract}
We introduce quantum circuits for simulations of multi-mode statevectors on 3D circuit Quantum Electrodynamics (cQED) processors, using matrix product state representations. The circuits are demonstrated as applied to simulations of molecular docking based on holographic Gaussian Boson Sampling (GBS), as illustrated for binding of a thiol-containing aryl sulfonamide ligand to the tumor necrosis factor-$\alpha$ converting enzyme receptor. We show that cQED devices with a modest number of modes could be employed to simulate multimode systems by re-purposing working modes through measurement and re-initialization. We anticipate a wide range of Gaussian boson sampling applications could be implemented on compact 3D cQED processors analogously, using the holographic approach. Simulations on qubit-based quantum computers could be implemented analogously, using circuits that represent continuous variables in terms of truncated expansions of Fock states.

\section{Introduction}
Graph theory plays a crucial role in computational chemistry, aiding in the modeling of molecules, chemical datasets, and reaction networks, as shown in recent studies.\cite{McDermott2021,Hashemi2022,Ratkiewicz2003,GD2008,Burch2019} It enables easier calculations in diverse areas from cheminformatics,\cite{Aldeghi2022} to quantum chemistry,\cite{Gutman1975,Gutman1986} and polymer chemistry.\cite{Mohapatra2022} Despite its utility, the complex nature of molecules and chemical databases often results in large graphs, challenging the capabilities of classical algorithms. For instance, several classes of problems require the computation of permanents, which is known to be a \#P-hard, escalating to \#P-complete for binary matrices.~\cite{VALIANT1979189} Such complexity places these computations beyond the reach of conventional methods.~\cite{Bezakova2008} Near-term quantum computers that implement Boson Sampling (BS)~\cite{Aaronson2011} offer a promising quantum solution. Here, we explore the application of BS to simulations of molecular docking, focusing on the possibility of sampling subgraphs describing the interactions between a molecule and a biological receptor using compact bosonic processors.

BS in its fundamental form involves the sampling of photons dispersed through a passive $N$-mode linear interferometer, as proposed by Aaronson and Arkhipov,~\cite{Aaronson2011} building upon the work by Troyansky and Tishby on quantum calculations of permanents and determinants.~\cite{troyansky1996quantum} In such experiments, the outcome distribution is determined by the permanent of the $N\times N$ matrix representing the transition probability amplitudes of the linear interferometer. As a result, BS enables the sampling of bosons from a distribution that would be challenging to simulate classically for large values of $N$, offering a promising approach to tackling problems previously considered intractable.

BS has been implemented on a diverse array of hardware platforms,~\cite{Motes2014,Thekkadath2022,Spagnolo2023,Wang2017} offering versatility and potential in applications of chemical relevance.~\cite{Wang2020,huh2015boson,Peropadre2016,Banchi2020,Quesada2022} Notably, BS has been applied to compute the vibronic spectra of triatomic molecules using 3D circuit Quantum Electrodynamics (cQED) processors,~\cite{Wang2020,huh2015boson} and to simulate molecular docking on photonic devices.~\cite{Peropadre2016,Banchi2020,Quesada2022} 
In fact, the initial BS sampling setup utilized optical photons. However, the efficient generation and detection of non-classical light states pose significant challenges.\cite{Peropadre2016,Wang2020} Here, we explore the use of microwave photons within the cQED architecture,~\cite{Blais2021} offering a promising alternative technology with unique advantages for implementing BS.

The cQED architecture stands at the forefront of quantum technology, featuring superconducting qubits based on Josephson junctions that are strongly coupled to the modes of superconducting microwave cavities. This setup facilitates efficient information exchange and enables universal quantum control over quantum states. Moreover, it supports quantum non-demolition (QND) measurements of photon numbers within the microwave cavities of the superconducting circuits. The precision in photon counting afforded by this architecture enhances the feasibility of implementing BS as evidenced by previous studies.~\cite{Wang2020}

In this paper, we explore the possibility of implementing BS simulations of molecular docking using 3D cQED processors. Molecular docking, an essential computational technique for drug design, predicts the binding configuration and most favorable interactions of a molecule and its receptor. This task is difficult for classical computers due to the need to perform an exhaustive search of possible molecular configurations. BS, with its ability to sample from a distribution defined by the permanent of the interaction matrix, holds promise for significantly enhancing the efficiency of molecular docking simulations, offering a novel approach to overcoming the limitations of traditional computational methods.

In this study, we focus on Gaussian boson sampling (GBS), a version of BS based on multimode Gaussian states.~\cite{hamilton2017gaussian,Bradler2018,Oh2024} We explore the possibility of simulating GBS using 3D cQED processors employing a holographic approach~\cite{he2017time,Motes2014,Foss2021}. This method is inspired by the successful utilization of GBS in conjunction with 3D cQED processors for simulations of molecules.\cite{Wang2020}

Our methodology begins by approximating the multimode statevectors as low-rank matrix product states~\cite{schon2005sequential,ran2020encoding} (MPS, also called tensor-trains~\cite{Oseledets11,lyu2022tensor,lyu2023tensor2}), followed by the variational parametrization of quantum circuits to represent these statevectors. The resulting circuits are executed holographically by re-purposing modes,\cite{he2017time,Motes2014,Foss2021} with the term "holographic" reflecting the dimension reduction benefit from the re-purposing algorithm, in echo with holography that constructs 3D image from 2D snapshots. This allows for hardware efficiency, enabling simulations of a few tens of modes with as few as 2-3 microwave modes in the cQED devices.

Our numerical simulations reveal that these cQED circuits could accurately simulate multimode Gaussian states, closely matching the benchmark Gaussian states of conventional GBS applied to molecular docking problems. Therefore, we anticipate these findings will pave the way for experimental investigations of compact 3D cQED devices into molecular docking and other subgraph isomorphism problems~\cite{Bradler2021} using, highlighting the potential for conducting realistic simulations with numerous modes on modestly sized cQED devices.

\section{Methodology}
\subsection{Molecular docking by Gaussian boson sampling}\label{sec:GBS}
\begin{figure*}
    \includegraphics[scale=0.75]{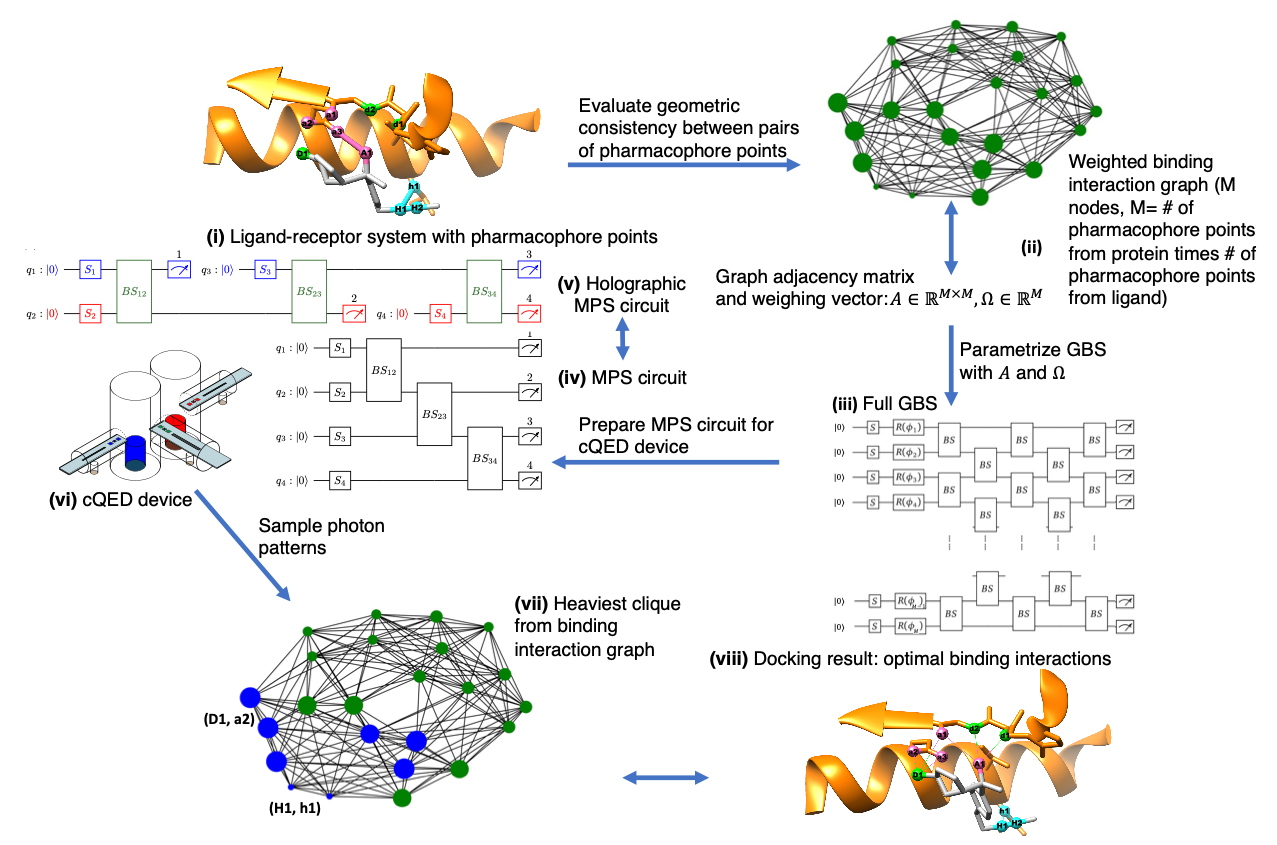}
    \caption{Schematic representation of the GBS approach implemented to investigate molecular docking due to ligand-receptor interaction (i). The interactions are represented by a graph where vertices correspond to pairs of interaction between pharmacophores in the ligand and in the receptor, while the edges connect pairs of contact that can be established simultaneously (ii). The graph adjacency matrix is used to define the covariance matrix of a multimode Gaussian state, generated by a passive linear interferometer of beam-splitters (iii). Sampling bosons from that distribution reveals the subsets of compatible pairs of interactions that can be established simultaneously. To sample from that distribution, initially a matrix product state (MPS) representation of the circuit is generated (iv), which can be implemented with a holographic approach (v) on a compact cQED device (vi). The observed photon patterns in the output modes indicate the heavier cliques (vii), from which optimal ligand-receptor interactions are established.}\label{fig:scheme}
\end{figure*}

Figure~\ref{fig:scheme} shows a schematic representation of the GBS methodology implemented to investigate molecular docking. Ligand-receptor interactions (Fig.~\ref{fig:scheme}, i) are established by ligand and receptor pharmacophores, as described in Sec.~\ref{sec:MDGSP}. The interaction graph (Fig.~\ref{fig:scheme}, ii)
describes the network of compatible pairs of interactions where the vertices correspond to pairs of interactions while the edges connect interactions that can be established simultaneously. Sec.~\ref{sec:MCGBS} outlines how the interaction graph is mapped into a GBS device to generate a multimode Gaussian statevector, with covariance parameterized by the adjacency matrix of the interaction graph (Fig.~\ref{fig:scheme}, iii).

Our simulations represent the statevector in MPS format (Fig.~\ref{fig:scheme}, iv), allowing for a holographic implementation (Fig.~\ref{fig:scheme}, v) using a cQED device (Fig.~\ref{fig:scheme}, vi). Sampling from that statevector distribution reveals subgraphs with maximum cliques (Fig.~\ref{fig:scheme}, vii) corresponding to subsets of interactions that can be established simultaneously (Fig.~\ref{fig:scheme}, viii). Sec.~\ref{sec:network} outlines experimental strategies for holographic implementations of GBS based on MPS networks with one and two layers of two-mode beam-splitters, parameterized to approximate the full GBS network. Sec.~\ref{sec:hardware} describes cQED devices based on bosonic 3D cQED devices for simulations of holographic MPS networks.

%

\subsection{Interaction graph}\label{sec:MDGSP}
Molecular docking algorithms predict favorable binding configurations of ligands (or drug-like molecules) as determined by interactions established by pharmacophores in the ligand with complementary pharmacophores in the target macromolecule (receptor). Configurations are ranked in terms of docking scores that give a rough estimate of relative binding affinities based on molecular descriptors.

\begin{figure*}
    \includegraphics[scale=0.6]{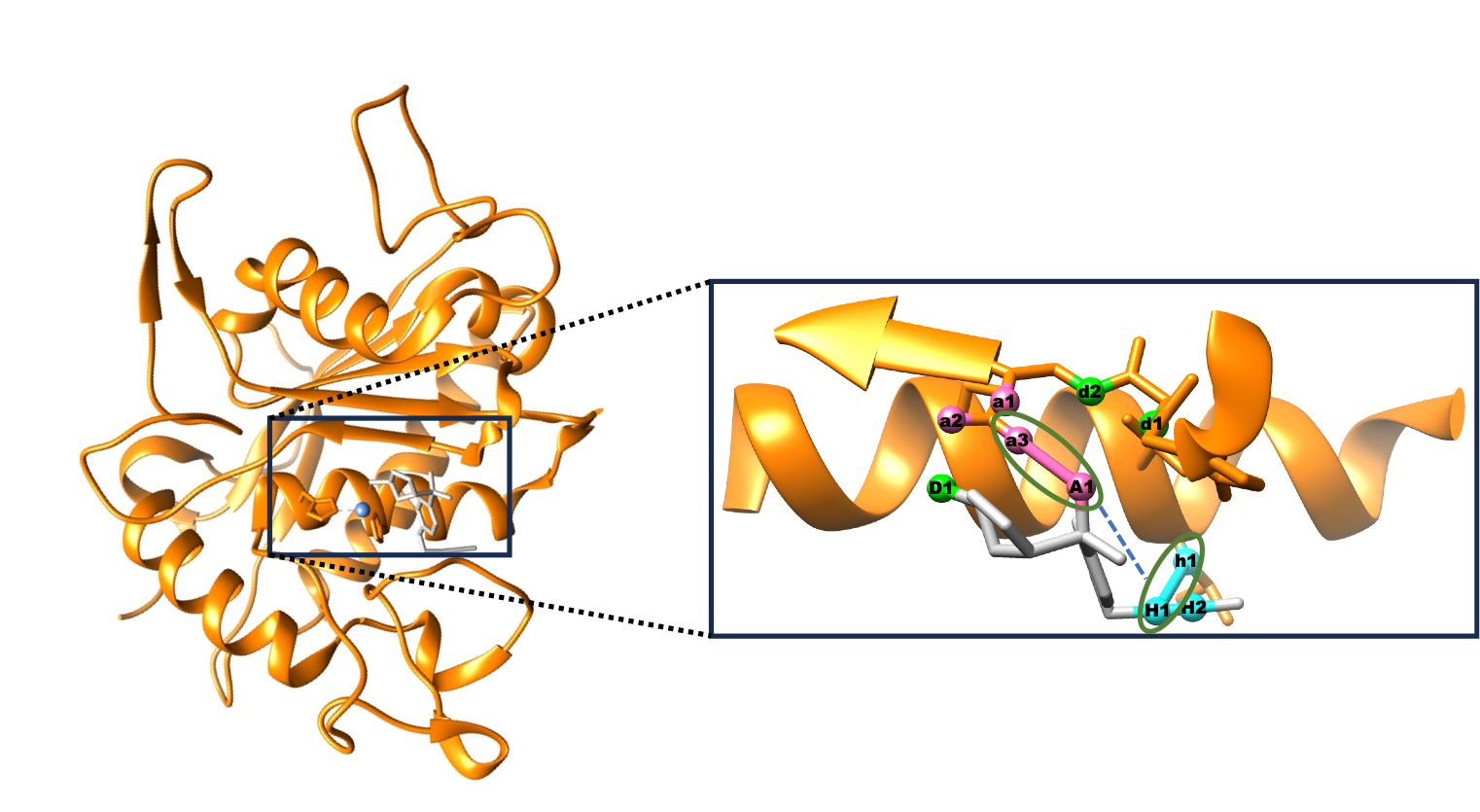}
    \includegraphics[scale=0.35]{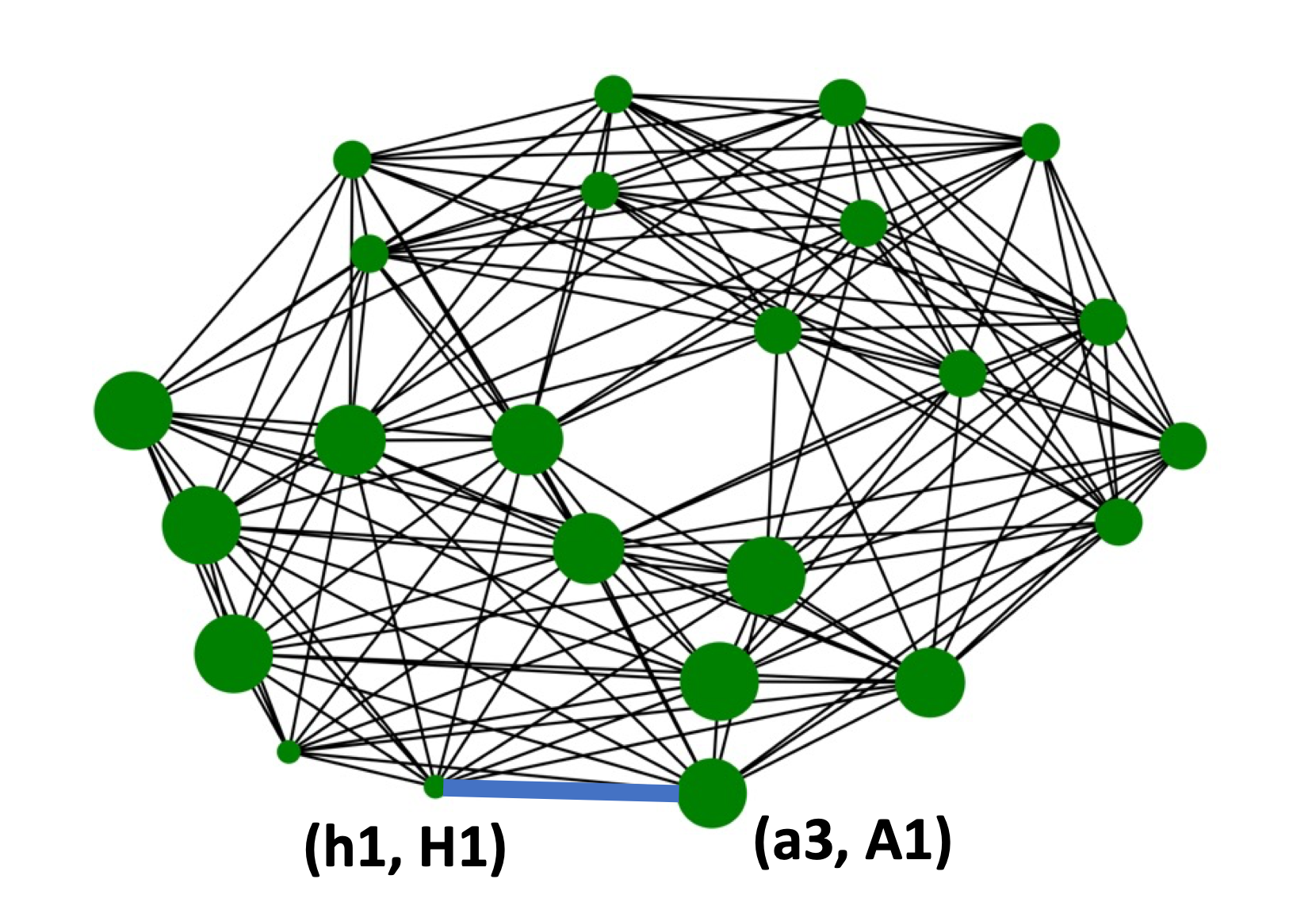}
    \caption{(Top left) Crystal structure of the tumor necrosis factor-$\alpha$ converting enzyme receptor (TACE, orange) with thiol-containing aryl sulfonamide ligand (AS, gray). (Top right) zoomed-in binding site with four pharmacophores in the ligand (A1, D1, H1, H2) and six in the receptor (a1, a2, a3, d1, d2, h1). Two pairs of binding interactions (a3, A1) and (h1, H1) can be concurrently established, as shown by blue dashed lines. (Bottom) Binding interaction graph of all possible pairwise ligand-receptor interactions. Larger nodes indicate heavier weights (stronger interactions).}\label{fig:igraph}
\end{figure*}

Reliable determination of docking scores for a series of ligands and favorable binding configurations is particularly valuable since it allows for rapid {\em in silico} screening of a large number of compounds. That process can quickly identify promising lead compounds for subsequent more accurate analysis, and discard unsuitable ligands. In the simplest approach, both the ligand and the receptor are approximated as rigid bodies, although methods that account for the inherent flexibility of the ligand and the receptor are available.~\cite{dias2008molecular} Favorable ligand orientations at the binding site can be revealed by using the isomorphous subgraph matching method, as implemented in the DOCK 4.0, FLOG, and SQ algorithms.~\cite{Bradler2021,kuhl1984combinatorial,miller1994flog,ewing1997critical,miller1999sq} In that formulation of the binding problem, the ligand-receptor interactions are represented by the so-called {\em interaction graph} $G(V, E)$, shown in Fig.~\ref{fig:igraph}. Here, the vertices $V$ represent ligand-receptor interactions established by pharmacophores,~\cite{yang2010pharmacophore} while the edges $E$ link pairs of interactions that can occur concurrently since the drug pharamacophores involved in these interactions are at the same distance apart as the corresponding pharmacophores in the receptor.

The interaction graph $G(V,E)$ is defined by its $M\times M$ adjacency matrix $A$, where $M>1$ is the total number of possible ligand-receptor interactions, with $A_{ij}=1$ if the interactions $i$ and $j$ are compatible (if both interactions can be established simultaneously), and $A_{ij}=0$, otherwise. To model the interaction strength between pharmacophores, we weight every vertex according the type of pharmacophores $i$ and $j$, biasing the strength of the intermolecular interactions with a pre-parameterized potential.~\cite{kitchen2004docking,poole2006knowledge,gohlke2001statistical} 
Having defined the adjacency, the problem of docking is reduced to the so-called {\em maximum clique problem},\cite{Bradler2021} which involves finding the largest fully connected subgraph within the graph, representing the largest set of compatible interactions as determined by the types of pharmacophore interactions and the inter-pharmacophore distances. 

\subsection{Maximum clique subgraph}\label{sec:MCGBS}
GBS addresses the maximum clique problem, as shown below, through preparation of a multimode Gaussian statevector whose covariance matrix is parameterized by the adjacency matrix of the ligand-receptor interaction graph. The statevector is generated from a vacuum state, using single-mode squeezing and a multimode linear interferometer. For simplicity, it is assumed that the mode count significantly exceeds the average photon number, resulting in either a single photon or no photon detection ($n_j\in\{0,1\}$) in each mode $j$.

\subsubsection{Multimode Gaussian of Bosonic Modes}
A Gaussian state of a system with $M$ bosonic modes, designated as $j$, with creation and annihilation operators $\hat{a}_j^{\dagger}$ and $\hat{a}_j$, can be described by the density matrix
\begin{align}\label{eq:mmg}
    \hat{\rho}
    =(2\pi)^{-M}|\det(\sigma)|^{-{1/2}}e^{-\frac{1}{2}(\hat{\xi}-d)^{\dagger}\sigma^{-1}(\hat{\xi}-d)},
\end{align}
where $\hat{\xi}=(\hat{a}_1,\dots,\hat{a}_M,\hat{a}_1^{\dagger},\dots,\hat{a}^\dagger_M)^T$ is a $2M$-component vector of operators. According to Eq.~(\ref{eq:mmg}), the Gaussian state is defined completely by its first and second statistical moments, $d$ and $\sigma$, respectively. The first moments
\begin{align}\label{eq:disp}
    d_j
    =\text{Tr}[\hat{\rho}\hat{\xi}_j]
\end{align}
describe the mode displacements, while the second moments
\begin{align}\label{eq:covar}
    \sigma_{ij}
    =\frac{1}{2}\text{Tr}[\hat{\rho}\{(\hat\xi_i-d_i),(\hat\xi_j-d_j)^\dagger\}]
\end{align}
define the $2M\times 2M$ covariance matrix $\sigma$ in terms of the anti-commutators
\begin{align}
    \{(\hat{\xi}_i-d_i),(\hat{\xi}_j-d_j)^\dagger\}=(\hat{\xi}_i-d_i)(\hat{\xi}_j-d_j)^\dagger+(\hat{\xi}_j-d_j)^\dagger(\hat{\xi}_i-d_i).
\end{align}
Measurements of the modes, when the system is described by the Gaussian statevector defined by Eq.~(\ref{eq:mmg}), report whether a photon is detected or not in each mode, as follows: $\bar{n}=(n_1,n_2,\dots,n_M)^T$, where $n_j\in\{0,1\}$ with $j=1,\dots,M$. In the realm of molecular docking, each mode $j$ symbolizes a distinct interaction between a pharmacophore of the drug and a pharmacophore of the receptor. Detecting a photon in a particular mode suggests that the corresponding interaction is established, whereas the absence of photons signifies a disruption of that interaction.

\subsubsection{Probability Distribution}
The probability distribution of outputs $\bar{n}$, obtained by measuring the multimode Gaussian in the Fock basis, is defined, as follows
\begin{align}\label{eq:Prn}
    \text{Pr}(\bar{n})
    =\text{Tr}[\hat{\rho}\hat{\bar{n}}],
\end{align}
where $\hat{\bar{n}}=\bigotimes_{j=1}^M\hat{n}_j$ is the tensor product of number state operators $\hat{n}_j=\vert n_j\rangle\langle n_j\vert$ corresponding to the probability of observing $n_j$ bosons in output mode $j$.
The right-hand side of Eq.~\eqref{eq:Prn} can be expanded by using the P-representation of operators, as follows (Appendix~\ref{sec:pfunc2})
\begin{align}\label{eq:expv_main}
    \text{Tr}\left[\bigotimes_{j=1}^M{\vert n_j\rangle\langle n_j\vert}\hat{\rho}\right]
    =\int P_{\bigotimes_{j=1}^M{\vert n_j\rangle\langle n_j\vert}}(\alpha)Q(\alpha)\,\mathrm{d}^2\alpha,
\end{align}
where
\begin{align}
    P_{\bigotimes_{j=1}^M{\vert n_j\rangle\langle n_j\vert}}(\alpha)
    =\prod_{j=1}^M\frac{e^{\vert\alpha_j\vert^2}}{n_j!}\frac{\partial^{2 n_j}}{\partial^{n_j}\alpha_j\partial^{n_j}\alpha_j^*}\delta^2{\alpha_j}.
\end{align}
$Q(\alpha)$, introduced by Eq.~(\ref{eq:expv_main}), is the Husimi function of the Gaussian state, defined as follows (Appendix~\ref{sec:pfunc2})
\begin{align}\label{eq:qfunc_main}
    Q(\alpha)
    =\pi^{-M}\langle\alpha\vert\hat{\rho}\vert\alpha\rangle
    =\frac{\pi^{-M}}{\sqrt{|\det(\sigma_Q)|}}e^{-\alpha^\dagger\sigma_Q^{-1}\alpha},
\end{align}
where $\alpha=(\alpha_1,\dots,\alpha_M,\alpha_1^*,\dots,\alpha_M^*)^T$ and $\sigma_Q=\sigma+I_{2M}/2$, with $I_{2M}$ being the $2M\times 2M$ identity matrix. The Husimi function obtained with Eq.~(\ref{eq:qfunc_main}), using the Gaussian density matrix introduced by Eq.~(\ref{eq:mmg}), can be substituted into Eq.~(\ref{eq:expv_main}) to obtain (Appendix~\ref{sec:output})
\begin{align}\label{eq:expv4_main}
    \text{Pr}(\bar{n})
    =\frac{1}{\sqrt{|\det(\sigma_Q)|}}\prod_{j=1}^M\frac{1}{n_j!}\frac{\partial^{2n_j}}{\partial^{n_j}\alpha_j\partial^{n_j}\alpha_j^*}\,e^{\frac{1}{2}\alpha^{T} K\alpha}\Big|_{\alpha_j=0},
\end{align}
where
\begin{align}\label{eq:adjj_main}
    K
    =X_{2M}(I_{2M}-\sigma_Q^{-1})
\end{align}
with $X_{2M}=
\begin{bmatrix}
    0 & I_M\\
    I_M & 0
\end{bmatrix}$.
Since $n_j\in\{0,1\}$, the multivariate derivatives in Eq.~\eqref{eq:expv4_main} can be easily evaluated by using the high-order chain rule given by the Faa di Bruno's formula (Appendix~\ref{sec:bruno}), as follows
\begin{align}\label{eq:Prn_pm}
    \text{Pr}(\bar{n})
    =\frac{1}{\bar{n}!\sqrt{|\det(\sigma_Q)|}}\sum_{\rho\in P_{2N}^2}\prod_{\{i,j\}\in\rho}K_{ij},
\end{align}
where $N=n_1+\dots+n_M$ is the total number of modes where a photon was detected according to $\bar{n}$, and $\bar{n}!=n_1!n_2!\dots n_M!$. Moreover, $P_{2N}^2$ is the set of partitions of the set $\{1,2,\dots,2N\}$ into subsets of two indices. The summation in Eq.~\eqref{eq:Prn_pm} is exactly the Hafnian of the $2N\times 2N$ submatrix of $K$ with rows and columns corresponding to the modes where photons were detected (\textit{i.e.}, modes $j$ with $n_j=1$).
Defining the submatrix of those modes as matrix $K_S$, we obtain
\begin{align}\label{eq:Prn_HafK}
    \text{Pr}(\bar{n})
    =\frac{1}{\bar{n}!\sqrt{|\det(\sigma_Q)|}}\text{Haf}(K_S).
\end{align}
To encode graph problems into Gaussian quantum states, we consider the matrix $K$ to be defined in terms of the graph adjacency matrix $A$, as follows~\cite{Bradler2018}
\begin{align}\label{eq:kaa}
    K
    =c(A\oplus A),
\end{align}
where the circled plus sign denotes the construction of a block-diagonal matrix from the component matrices. Furthermore, $0<c<\lambda_1^{-1}$ is a positive rescaling parameter, with $\lambda_1>0$ the maximum eigenvalue of $A$. Note that for a undirected graph without self-connection, $A$ is a real symmetric matrix with all diagonal elements being $0$. Since $\text{Tr}[A]=0$ implies that the eigenvalues must be either all 0 or containing both positive and negative numbers, and the only real symmetric matrix that contains all zero eigenvalues is the zero matrix, the largest eigenvalue of $A$ must be positive for a non-zero $A$. In particular, the scaling factor ensures that $K$ corresponds to a valid covariance matrix $\sigma$ as in Eq.~\eqref{eq:adjj_main}, resulting in a probability distribution $\text{Pr}(\bar{n})$ that is bounded between zero and one. In Sec.~\ref{sec:circuit} we show that an $M$-mode Gaussian state with such a $K$ can always be prepared by a programmed optical network.

Substituting Eq.~\eqref{eq:kaa} into Eq.~\eqref{eq:Prn_HafK}, we obtain
\begin{align}\label{eq:Prn_A}
    \begin{split}
        \text{Pr}(\bar{n})
        &=\frac{c^N}{\bar{n}!\sqrt{|\det(\sigma_Q)|}}\text{Haf}(A_S\oplus A_S)\\
        &=\frac{c^N}{\bar{n}!\sqrt{|\det(\sigma_Q)|}}\text{Haf}(A_S)^2,
    \end{split}
\end{align}
providing the probability distribution of outputs $\bar{n}$ in terms of the Hafnian of $A_S$ (with $A_S$ the submatrix of $A$ defined by the intersection of rows and columns of $A$ corresponding to the modes where photons were detected).

\subsubsection{Binary Graphs}
In the simplest formulation of a binary graph, $A$ is the $M\times M$ real and symmetric adjacency matrix with $A_{ij}\in\{0,1\}$. $A_{ij}=1$ denotes that nodes $i$ and $j$ share an edge, while $A_{ij}=0$ denotes that nodes $i$ and $j$ are not connected. In the context of binding interaction graphs that are of our primary interest, a node denotes a binding interaction between a pair of pharmacophore points, and connection between two nodes indicate the two binding interactions can be established simultaneously, while no connection indicates that the two binding interactions cannot be concurrently established because by establishing one interaction the other one is disrupted. This binary approach can be generalized to account for the strength of the connections, using instead a weighted binding interaction graph as described later in Sec.~\ref{sec:wei}.

It is important to note that each term in the sum of the Hafnian in Eq.~\eqref{eq:Prn_A} represents a {\em perfect matching} where each node of the subgraph is connected to one (and only one) other node. As a result, the Hafnian gives the count of perfect matchings, and thus reveals the number of possible ways the drug pharamacophores can concurrently establish interactions with pharmacophores in the receptor. As per Eq.~\eqref{eq:Prn_A}, subgraphs with larger Hafnian are sampled with higher probability.

Gaussian states are prepared with covariance $\sigma$, obtained by inverting Eq.~(\ref{eq:adjj_main}), as follows
\begin{align}\label{eq:iadjj_main}
    \sigma
    =\left(I_{2M}-X_{2M}K\right)^{-1}-\frac{I_{2M}}{2},
\end{align}
with $K$ defined according to Eq.~(\ref{eq:kaa}). This is achieved by using a quantum circuit that integrates single-mode squeezing gates and a multimode linear interferometer, as detailed in Sec.~\ref{sec:circuit}. Sampling from the statevector generated by that circuit enable us to effectively identifying the subgraphs with larger Hafnian within the interaction graph, revealing the capability of the drug to establish concurrent interactions with the receptor.

\subsubsection{Weighted Graphs}\label{sec:wei}
Extending the method to accommodate weighted graphs, beyond binary adjacency matrices $A$ (with $A_{ij}\in\{0,1\}$) is straightforward. This requires a weighting vector $\Omega$, where each element $\Omega_{ii}=c(1+w_i)$ is defined by a weight $w_i$, assigned to the $i^{th}$ node. This design of weighting vector ensures that subgraphs with a larger weight are favoured during sampling.\cite{Banchi2020} In the context of molecular docking, these weights are obtained from a knowledge-based potential, where a heavier weight $w_i$ corresponds to a stronger binding interaction.\cite{Bradler2021} For instance, if the $i^{th}$ node corresponds to a hydrogen bond donor-acceptor interaction, $w_i$ would be larger than nodes representing weaker interactions (e.g., hydrophobic contacts). The weighting vector $\Omega$ is written as an diagonal matrix ($\text{diag}(\Omega)\rightarrow\Omega$) and applied on both sides of $A$, to generate a weighted graph adjacency matrix, as follows
\begin{align}
    A\mapsto\Omega A\Omega,
\end{align}
generates a photon distribution defined by both $A$ and $\Omega$, as follows
\begin{align}\label{eq:whaf}
    \begin{split}
        \text{Pr}(\bar{n})
        &=\frac{c^N}{\bar{n}!\sqrt{|\det(\sigma_Q)|}}[\text{Haf}(\Omega_S A_S\Omega_S)]^2\\
        &=\frac{c^N}{\bar{n}!\sqrt{|\det(\sigma_Q)|}}\left[\sum_{\rho\in P_{N}^2}\prod_{\{i,j\}\in\rho}\Omega_{ii}A_{ij}\Omega_{jj}\right]^2\\
        &=\frac{c^N}{\bar{n}!\sqrt{|\det(\sigma_Q)|}}\left[\det(\Omega)\sum_{\rho\in P_{N}^2}\prod_{\{i,j\}\in\rho}A_{ij}\right]^2\\
        &=\frac{c^N}{\bar{n}!\sqrt{|\det(\sigma_Q)|}}\left[\det(\Omega_S)\text{Haf}(A_S)\right]^2,
    \end{split}
\end{align}
where $\Omega_S$ is the submatrix of $\Omega$ corresponding to the modes registering photons (the modes that also define $A_S$). According to Eq.~(\ref{eq:whaf}), the resulting GBS with rescaled adjacency matrices has higher probability of sampling cliques with a large number of strongly interacting pharmacophores.

\subsubsection{Quantum Circuit}\label{sec:circuit}
Here, we show how to build a quantum circuit of $M$ modes that generates the desired multimode Gaussian state with covariance defined by Eq.~(\ref{eq:iadjj_main}) with $K=c(A\oplus A)$, as necessary for GBS. In particular, we show that such a state can be obtained by passing a vacuum Gaussian state through an optical network composed of $M$ single-mode squeezers and an $M$-mode interferometer, both parameterized according to $A$.
Firstly, we show how the single-mode squeezers and $M$-mode interferometer transform the $M$-mode vacuum state by preserving its Gaussian shape but changing its covariance. Then, we show how the squeezers and interferometer are parameterized according to $A$.

The single mode vacuum state $|0\rangle$ is the eigenstate of $\hat{a}$ with eigenvalue 0:
\begin{align}\label{eq:vac}
    \hat{a}|0\rangle
    =0.
\end{align}
Therefore,
\begin{align}\label{eq:vac_eval}
    \begin{split}
        \langle\hat{a}\rangle_{\text{vac}}
        &=\text{Tr}[|0\rangle\langle 0|\hat{a}]
        =0,\\
        \langle\hat{a}^\dagger\rangle_{\text{vac}}
        &=\text{Tr}[|0\rangle\langle 0|\hat{a}^\dagger]
        =0,\\
        \langle\hat{a}\hat{a}\rangle_{\text{vac}}
        &=\text{Tr}[|0\rangle\langle 0|\hat{a}\hat{a}]
        =0,\\
        \langle\hat{a}\hat{a}^\dagger\rangle_{\text{vac}}
        &=\text{Tr}[|0\rangle\langle 0|\hat{a}\hat{a}^\dagger]
        =1,\\
        \langle\hat{a}^\dagger\hat{a}\rangle_{\text{vac}}
        &=\text{Tr}[|0\rangle\langle 0|\hat{a}^\dagger\hat{a}]
        =0,\\
        \langle\hat{a}^\dagger\hat{a}^\dagger\rangle_{\text{vac}}
        &=\text{Tr}[|0\rangle\langle 0|\hat{a}^\dagger\hat{a}^\dagger]
        =0,
    \end{split}
\end{align}
where the subscripts indicate that the expectation values $\langle\bullet\rangle$ are evaluated with $\hat{\rho}=|0\rangle\langle 0|$. Substituting Eq.~\eqref{eq:vac_eval} into Eq.~\eqref{eq:covar}, we obtain the covariance matrix of the single mode vacuum state
\begin{align}
    \sigma_{\text{vac}}
    =\frac{1}{2}
    \begin{bmatrix}
        1 & 0\\
        0 & 1
    \end{bmatrix}.
\end{align}
The single-mode squeezing operation is defined, as follows
\begin{align}
    \hat{S}(r)
    =e^{(-r(\hat{a})^2+r(\hat{a}^{\dagger})^2)/2},
\end{align}
where $r\in\mathbb{R}$ is real-valued and is referred to as the squeezing parameter. Appendix~\ref{sec:sque-cov} (Eq.~\eqref{eq:sque-bogo}) shows that in the Heisenberg picture, the action of the squeezing operation transforms the annihilation and creation operators, as follows
\begin{align}\label{eq:sque-aadag}
    \begin{split}
        \hat{a}^{'}
        &=\hat{S}(r)^\dagger\hat{a}\hat{S}(r)
        =\text{cosh}(r)\hat{a}+\text{sinh}(r)\hat{a}^\dagger,\\
        (\hat{a}')^{\dagger}
        &=\hat{S}(r)^\dagger\hat{a}^\dagger\hat{S}(r)
        =\text{cosh}(r)\hat{a}^\dagger+\text{sinh}(r)\hat{a}.
    \end{split}
\end{align}
Substituting Eq.~\eqref{eq:sque-aadag} into Eq.~\eqref{eq:covar}, we find that the single mode squeezed state is again Gaussian with the following covariance matrix (Eq.~\eqref{eq:squeez_cov})
\begin{align}\label{eq:sque_cov}
    \mathbf{\sigma}'(r)
    =\frac{1}{2}
    \begin{bmatrix}
        \text{cosh}^2(r)+\text{sinh}^2(r) & 2\text{cosh}(r)\text{sinh}(r)\\
        2\text{cosh}(r)\text{sinh}(r) & \text{cosh}^2(r)+\text{sinh}^2(r)
    \end{bmatrix}.
\end{align}
Generalizing to $M>1$ modes, we have
\begin{align}\label{eq:M_sque_cov}
    \sigma_{\text{sque}}
    =\frac{1}{2}
    \begin{bmatrix}
        \bigoplus_{j=1}^M\text{cosh}^2(r_j)+\text{sinh}^2(r_j) & \bigoplus_{j=1}^M 2\text{cosh}(r_j)\text{sinh}(r_j)\\
        \bigoplus_{j=1}^M 2\text{cosh}(r_j)\text{sinh}(r_j)& \bigoplus_{j=1}^M\text{cosh}^2(r_j)+\text{sinh}^2(r_j)\\
    \end{bmatrix}.
\end{align}
Next, we evaluate the effect of the $M$-mode interferometer on the squeezed state covariance matrix by starting from the simplest 2-mode case and then obtaining the $M$-mode case.

The smallest interferometer is the 2-mode beam-splitter. It transforms a two-mode state according to the following operator
\begin{align}
    \hat{B}(\theta)
    =e^{\theta(\hat{a}_1^\dagger\hat{a}_2-\hat{a}_1\hat{a}_2^\dagger)},
\end{align}
where $\theta\in[0,2\pi]$ is a given angle. In the Heisenberg picture, $\hat{B}(\theta)$ transforms the annihilation operators, as follows
\begin{align}
    \begin{split}
        \hat{a}_1^{'}
        &=B(\theta)^\dagger\hat{a}_1B(\theta)
        =\text{cos}(\theta)\hat{a}_1+\text{sin}(\theta)\hat{a}_2\\
        \hat{a}_2^{'}
        &=B(\theta)^\dagger\hat{a}_2B(\theta)
        =\text{cos}(\theta)\hat{a}_2-\text{sin}(\theta)\hat{a}_1.
    \end{split}
\end{align}
Hence, for $k=1,2$ we obtain
\begin{align}
    \begin{split}
        \hat{a}_k^{'}
        &=\sum_{j=1}^2U_{kj}\hat{a}_j,\\
        (\hat{a}_k')^{\dagger}
        &=\sum_{j=1}^2 U_{kj}^*\hat{a}_j^\dagger,
    \end{split}
\end{align}
where $U=U(\theta)=
\begin{bmatrix}
    \text{cos}(\theta) & \text{sin}(\theta)\\
    -\text{sin}(\theta) & \text{cos}(\theta)
\end{bmatrix}$
is a unitary transformation. Additionally, when a phase-shifter is placed after the beam-splitter such that $\hat{a}_1\mapsto e^{i\phi_1}\hat{a}_1$ for $\phi_1\in[0,2\pi]$, the resulting unitary transformation becomes $U=U(\theta,\phi_1)=
\begin{bmatrix}
    \text{cos}(\theta)e^{i\phi_1} & \text{sin}(\theta)e^{i\phi_1}\\
    -\text{sin}(\theta) & \text{cos}(\theta)
\end{bmatrix}$.
Note that any $2\times 2$ unitary matrix $U$ can be implemented like this using a beam-splitter and a phase-shifter with a suitable choice of values for $\theta$ and $\phi_1$.~\cite{reck1994experimental} 
The generalization to $M$ modes is straightforward and involves an $M$-mode interferometer, as follows
\begin{align}\label{eq:rot_aadag}
    \begin{split}
        \hat{a}_k^{'}
        &=\sum_j^MU_{kj}\hat{a}_j,\\
        (\hat{a}_k')^{\dagger}
        &=\sum_j^MU_{kj}^*\hat{a}_j^\dagger,
    \end{split}
\end{align}
where $U$ is now an $M\times M$ unitary matrix. Substituting Eq.~\eqref{eq:rot_aadag} into Eq.~\eqref{eq:covar}, we find that the $M$-mode interferometer $U$ transforms an arbitrary covariance matrix $\tilde{\sigma}$, as follows (Eq.~\eqref{eq:N_rot_cov})
\begin{align}
    \mathbf{\sigma}_{\text{rot}}
    =
    \begin{bmatrix}
        U & 0\\
        0 & U^*
    \end{bmatrix}
    \tilde{\sigma}
    \begin{bmatrix}
        U^* & 0\\
        0 & U
    \end{bmatrix}^T.
\end{align}
Therefore, the outgoing Gaussian state obtained by rotating an $M$-mode squeezed vacuum state is described by the following covariance matrix
\begin{align}\label{eq:n-rot-squeezed-cov_main}
    \sigma_{\text{out}}
    =
    \begin{bmatrix}
        U & 0\\
        0 & U^*
    \end{bmatrix}
    \sigma_{\text{sque}}
    \begin{bmatrix}
        U^* & 0\\
        0 & U
    \end{bmatrix}^T.
\end{align}
With the general covariance matrix defined, we now explain how the parameters of the squeezers and interferometer are obtained to ensure that $\sigma_{\text{out}}$ corresponds to Eq.~\eqref{eq:iadjj_main}, with kernel $K$ defined by Eq.~\eqref{eq:kaa}

According to Takagi's matrix factorization, the symmetric matrix $A$ can be decomposed, as follows
\begin{align}
    A
    =U\Sigma U^T
    =U\bigg(\frac{1}{c}\bigoplus_{j=1}^M\text{tanh}(r_j)\bigg)U^T,
\end{align}
where $\Sigma$ is the diagonal matrix of singular values of $A$, and $0<c<\lambda_1^{-1}$ is a rescaling parameter, with $\lambda_1>0$ the maximum eigenvalue of $A$. The rescaling ensures that every singular value can be represented as $\text{tanh}(r_j)/c$ (note that $\vert\text{tanh}(r)\vert<1$ for all $r\in\mathbb{R}$). The unitary matrix $U$ is used to parameterize the $M$-mode interferometer. The resulting covariance defined by Eq.~(\ref{eq:n-rot-squeezed-cov_main}) is consistent with the kernel $K$, introduced by Eq.~\eqref{eq:kaa}, as shown in Appendix~\ref{sec:K-N-rot} where Eq.~\eqref{eq:kaa} is obtained by substituting Eq.~\eqref{eq:n-rot-squeezed-cov_main} into Eq.~\eqref{eq:adjj_main}.

Section~\ref{sec:network} introduces an efficient implementation of the multimode Gaussian, with hardware efficiency, by first squeezing each mode and then implementing $U$ in MPS format by using an $M$-mode passive interferometer. Measurement of the modes generates a probability distribution of photon outcomes as defined by Eq.~(\ref{eq:whaf}). We note that the theoretical analyses in this subsection neglects errors in the squeezed states, which could potentially hinder the performance of the quantum device.\cite{Drummond2022} More detailed investigation is the subject of follow up researches of this initial theoretical proposal.

\subsection{Holographic implementation}\label{sec:network}
Figures~\ref{fig:MPS1} and \ref{fig:MPS2} illustrate the implementation of circuits corresponding to the state of a system of 4 modes, such as a multimode Gaussian state, in MPS format. Note that each beam-splitter (BS) couples two adjacent bosonic modes.

\subsubsection{MPS Circuits}\label{sec:circuits}
\begin{figure}[H]
    \includegraphics[scale=.90]{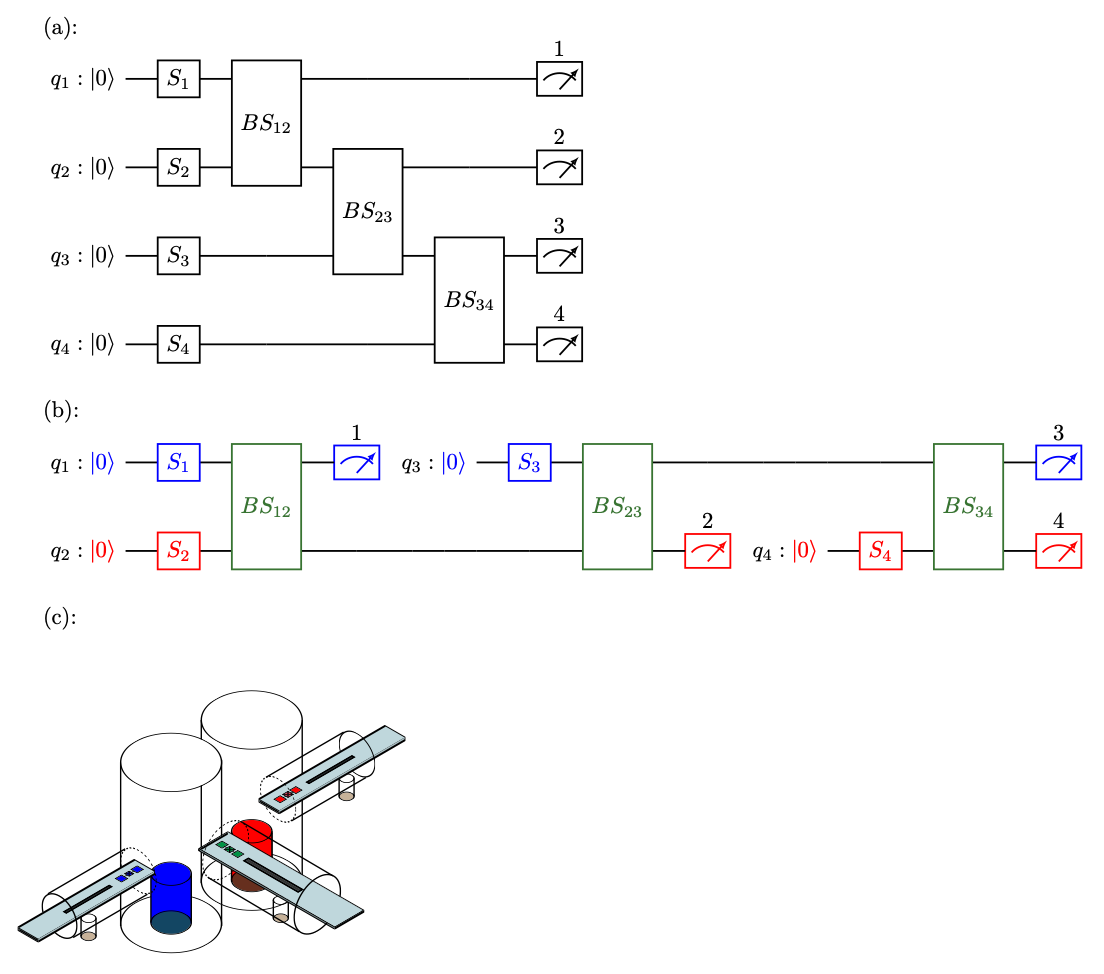}
    \caption{(a): Quantum circuit for a system with 4 modes, implemented by squeezing gates and one layer of beam-splitters. (b): Holographic implementation of the circuit shown in panel (a) by re-purposing two modes. The upper wire (blue) and lower wire (red) correspond to the two cavities of the same colors in (c), coupled by a transmon as beam-splitter (olive green). (c): 3D cQED device of two modes implemented with microwave cavity resonators coupled by a transmon.}\label{fig:MPS1}
\end{figure}

Figure~\ref{fig:MPS1}(a) shows the MPS circuit with one layer of beam-splitters, inspired by a circuit previously proposed for moderately entangled quantum states.~\cite{schon2005sequential} Fig.~\ref{fig:MPS1}(c) shows a compact 3D cQED device that could implement the circuit with hardware efficiency using the so-called holographic quantum computing approach shown in Fig.~\ref{fig:MPS1}(b). Note that the circuits shown in Figs.~\ref{fig:MPS1} (a) and (b) are equivalent. The only difference is that the circuit in (b) re-purposes the modes after measurement. Mode 1 ($q_1$) is measured after squeezing and coupling with mode 2 by implementing the gates $S_1$, $S_2$, and $BS_{12}$. After that measurement, mode 1 can be repurposed for representing mode 3 ($q_3$) by first squeezing with $S_3$ and coupling it with mode 2 according to $BS_{23}$. Mode 2 is measured and then re-purposed as mode 4. After applying $S_4$, $BS_{34}$, mode 3 and 4 are measured completing the implementation of the circuit with 4 modes, using a 2-mode quantum device. Circuits with more modes could be implemented analogously with the 2-mode device by iterative squeezing, coupling and measuring modes. Figure~\ref{fig:MPS1}(a) and (b) also show that the gate cost of our holographic algorithm is $O(m)$ for an $m$-mode Gaussian state.

\begin{figure}[H]
    \includegraphics[scale=1.0]{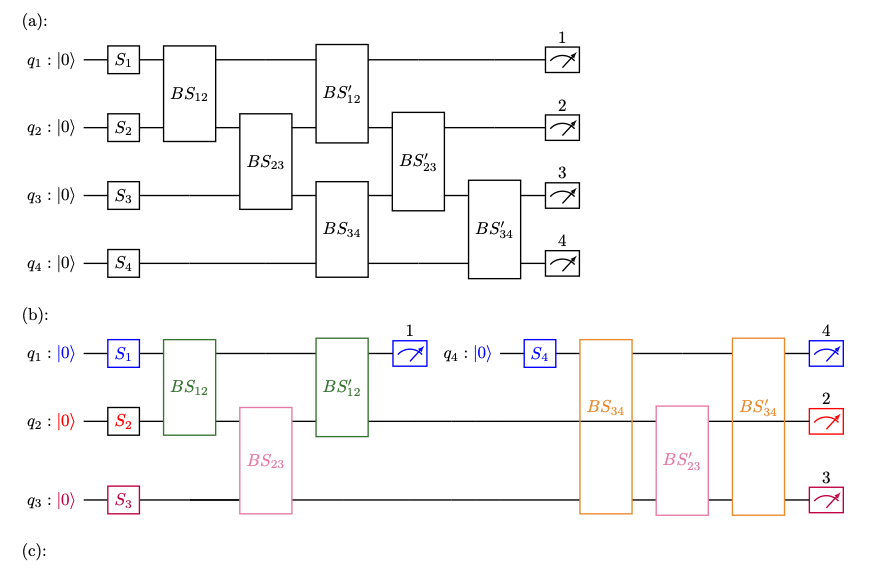}
    \includegraphics[scale=0.8]{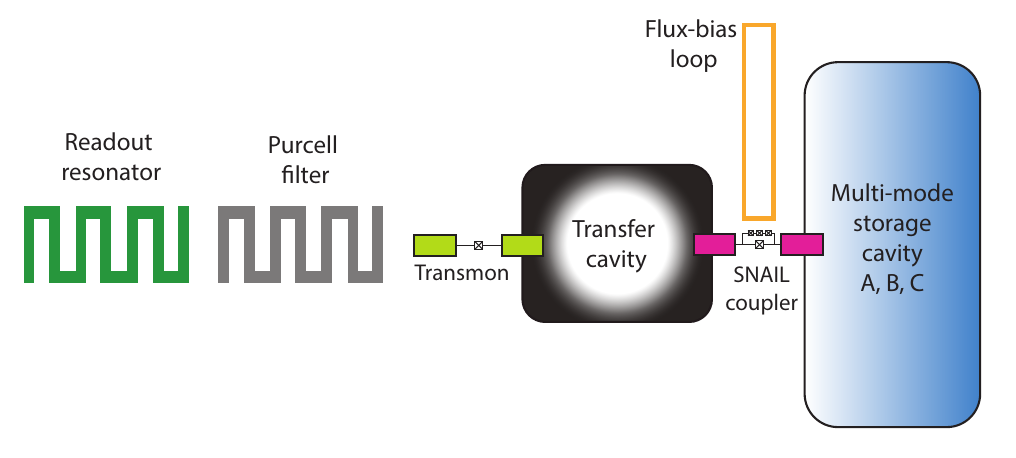}
    \caption{(a): Quantum circuit for a system with 4 modes, implemented by squeezing gates and two layers of beam-splitters. (b) Holographic implementation of the circuit shown in panel (a) by re-purposing three modes. Gates with the same color are simultaneously applied. (c) Multi-mode 3D cQED device implemented with microwave cavity resonators coupled by a transmon implemented as a quantum router.}\label{fig:MPS2}
\end{figure}

States with more entanglement can also be implemented with hardware efficiency by employing a few more modes. As an example, Fig.~\ref{fig:MPS2}(a) shows a 4-mode circuit with a higher level of entanglement established by a second layer of beam-splitters. Here, we implement the circuit by employing a 3-mode device. Figure~\ref{fig:MPS2}(b) shows the holographic implementation starting with gates $S_1$, $S_2$, $S_3$, $BS_{12}$, $BS_{23}$, $BS_{12}'$, that could be implemented with the device illustrated in Fig.~\ref{fig:MPS2}(c). Mode 1 would be measured first and re-purposed for representing mode 4. After this resetting, gates $S_4$, $BS_{34}$, $BS_{23}'$, $BS_{34}'$ are implemented, and modes 2-4 are finally measured. The resulting sampling is equivalent to that of the conventional GBS experiment.
%

\subsubsection{Parametrization}
In this section, we explain how we parameterize the beam-splitters employed in the MPS circuits introduced in Sec.~\ref{sec:circuits}, including the quantum circuits shown in Figs.~\ref{fig:MPS1} and~\ref{fig:MPS2} with one and two layers of beam-splitters. As discussed in Sec.~\ref{sec:results}, the resulting parametrization allows for efficient and accurate GBS simulations based on arbitrary adjacency matrices.

First, we obtain the covariance matrix of the reference state $\sigma_{\text{ref}}$ as well as the single-mode squeezing parameters from the adjacency matrix $A$, as described in Sec.~\ref{sec:circuit}. Next, we prepare the single-mode squeezed states from vacuum and fed them into the MPS-based network with variational beam-splitting and phase-shifting parameters $(\mathbf{\theta}=(\theta_1,\dots,\theta_K),\mathbf{\phi}=(\phi_1,\dots,\phi_K))$ for a network with a total of $K$ beam-splitters. These parameters are optimized by minimization of the Frobenius distance $d_{\text{tmp}}(\theta,\phi)=||\sigma_{\text{tmp}}(\theta,\phi)-\sigma_{\text{ref}}||_F$ using a standard conjugate gradient-descent optimizer, which has the computational complexity of O($dm^4$) for an $m$-mode covariance matrix, with $d$ is the number of iterations. The parameter set $\theta_{\text{fin}},\phi_{\text{fin}}$ corresponding to the minimal distance is then used as parameters for the quantum circuits.

\subsection{Bosonic cQED hardware setup for holographic GBS}\label{sec:hardware}
The experimental requirements for implementing the holographic GBS include a high-coherence multi-mode cavity system (with at least 2 storage modes), pair-wise beam-splitter operations with programmable phases and splitting ratios, and efficient readout, reset, and squeezing of individual cavity states. We emphasize that all of these requirements are already met simultaneously in existing hardware demonstrations, although design optimization is needed. For example, the two-mode device used to demonstrate bosonic two-qubit gates in Ref.~\citenum{gao_entanglement_2019}, as reproduced in Fig.~\ref{fig:MPS1}(c), can already be used to implement GBS of arbitrarily number of modes in principle, using the circuits shown in Fig.~\ref{fig:MPS1}(b). More recent demonstration of a quantum router connecting 4 storage modes~\cite{zhou2023realizing} provides all the tools necessary to implement the holographic GBS routine with added beam-splitter layers such as those shown in Fig.~\ref{fig:MPS2}(b). In this section, we briefly review the techniques necessary to physically implement the required bosonic operations, discuss the limitations of existing hardware, and introduce a minimal construction of 3D circuit QED device tailored for holographic GBS.

Simply speaking, the key figure of merit that characterizes the performance of a device hardware for the proposed GBS protocol is the ratio of cavity coherence times and the operational cycle time. This ratio sets an upper bound on the executable circuit depth or effective number of modes that can be computed. The current state of the art in multi-mode cavity QED has led to quantum control of over 10 cavity modes with millisecond lifetimes~\cite{chakram_seamless_2021}. The coaxial stub geometry, which is more widely adopted, consistently exhibits coherence times at the millisecond level. It also offers spatial separation of the modes, facilitating individual control~\cite{gao_entanglement_2019}. In principle, seamless designs of superconducting cavities, leveraging low surface-to-volume ratio and materials processing technologies borrowed from particle accelerators, can achieve coherence times on the order of seconds.~\cite{romanenko_three-dimensional_2020} This represents a significant potential for enhancing system performance by several orders-of-magnitude.

Programmable two-mode beam-splitter and single-mode squeezing gates can be achieved with either the four-wave mixing or the three-wave mixing process using external microwave pumps in 3D circuit QED. In either case, the external pump, applied under the appropriate frequency matching condition, can activate a rotating-frame Hamiltonian of the form $\hat{a}^\dagger\hat{b}e^{i\theta_i}+\hat{a}\hat{b}^\dagger e^{-i\theta_i}$ for photon conversion between $\hat{a}$ and $\hat{b}$ modes or $\hat{a}^{\dagger2} e^{i\phi_i}+\hat{a}^2 e^{-i\phi_i}$ for squeezing drives in mode $\hat{a}$ for the $i^{th}$ beamsplitter. The phase of the pump tone controls the beam-splitting phase $\theta_i$ or the squeezing phase $\phi_i$, while the amplitude and length of the external pump controls the splitting ratio or squeezing ratio. The four-wave mixing process is ubiquitously available in standard circuit QED hardware systems today, requiring only a fixed-frequency transmon ancilla, which generally allows for gates on the order of several $\mu$s~\cite{gao_programmable_2018}. Much faster gates on 100 ns scale or shorter can be implemented with parametric charge or flux drives~\cite{chapman_high--off-ratio_2023, lu_high-fidelity_2023} using novel Josephson ancilla circuitry that supports three-wave mixing processes. These ancillae including, for example, SNAIL (Superconducting Nonlinear Asymmetric Inductive eLement)~\cite{zhou2023realizing, frattini_3-wave_2017, Hillmann2020, Eriksson2024}, RF SQUID, ATS~\cite{lescanne_exponential_2020} have been under intensive development recently. Their integration with high-coherence 3D systems has led to record beam-splitter fidelity in the range of 99.9\%-99.99\%~\cite{chapman_high--off-ratio_2023, lu_high-fidelity_2023}.

Measurement and reset of cavities can use transmons ancillae and their low-Q readout resonators. A transmon qubit dispersively coupled to both a storage cavity and a readout resonator has been a well-developed tool to analyze the storage photon number using Ramsey sequences or selective transmon excitations~\cite{vlastakis_deterministically_2013}. If the task is to distinguish between $\ket{0}$ and $\ket{1}$ for the cavity, a Ramsey-like protocol for photon-number parity measurement is preferable for its faster speed, which should allow cavity measurements within 1 $\mu$s for typical coupling parameters. Cavity-state reset for arbitrary initial photon number can employ a beam-splitter between the cavity and a linear low-Q mode (such as the readout resonators in Fig.~\ref{fig:MPS1}(c). To achieve a reasonable operation speed ($\mu$s or less), however, this requires a dedicated parametric coupler capable of strong 3-wave mixing (e.g.~a SNAIL). Assuming the expected number of cavity photons is low, an alternative approach involves resetting specific number states to vacuum through intermediary transmon states. Specifically, it is possible to convert a single-photon excitation in the storage cavity into double excitations in the transmon using a four-wave mixing drive. Subsequently, resetting the transmon should allow the resetting of the cavity $\ket{1}$ state within 1 $\mu$s.

In constructing a full device capable of the proposed holographic GBS, a key consideration is the ability to measure and reset one cavity mode without affecting the other cavity modes. Therefore, a transmon ancilla with simultaneous dispersive coupling to multiple storage modes for their readout, such as in Ref.~\citenum{chakram_seamless_2021} is undesirable. Instead of designing spatially separated of storage cavity modes with independent transmon ancillas as in Refs.~\citenum{gao_programmable_2018, zhou2023realizing}, we propose hardware-efficient setup that satisfies all the requirements as in Fig.~\ref{fig:MPS2}(c). The storage cavity module containing multiple modes is coupled to a 3-wave mixing coupler (e.g. SNAIL) operated under the condition to minimize the 4th order nonlinearity of the storage modes (self Kerr and cross Kerr). The 3-wave coupler is further coupled to another high-Q transfer cavity with a transmon ancilla. The 3-wave coupler allows fast beam-splitter operations between storage modes as well as swapping between any storage mode and the transfer cavity (typically prepared in vacuum) for readout and initialization. The number of photons in the transfer cavity can be analyzed by leveraging the dispersive interaction with the transmon. Additionally, the reset of the transfer cavity can be carried out using transmon-mediated sideband drives, which are conditioned on the cavity states. Utilizing the transfer cavity not only protects the storage modes from spurious fourth-order nonlinearities but also protects the low-Q components of the circuits, which are required for fast measurement and reset.

\section{Results and Discussion}\label{sec:results}
We first demonstrate the capabilities of the MPS-based holographic approach as applied to GBS for solving the problem of molecular docking (Sec.~\ref{sec:mdock}). Next, we analyze the scalability of the approach as applied to random adjacency matrices (Sec.~\ref{sec:rad}). 

\subsection{Molecular docking mapped as a graph search problem}\label{sec:mdock}
In this section, we apply the MPS-based holographic approach to solve the molecular docking problem which involves finding the optimal binding mode of a small drug molecule bound to a target biological receptor. Specifically, we focus on the benchmark model system of a thiol-containing aryl sulfonamide compound (AS) bound to the tumor necrosis factor-$\alpha$ converting enzyme (TACE), shown in Fig.~\ref{fig:igraph}, that allows for direct comparisons of our MPS-based GBS approach to full GBS simulations.~\cite{Banchi2020} Binding of AS to TACE is determined by hydrogen-bonds and hydrophobic contacts established by 6 pharmacophores in the TACE active site and 4 pharmacophores in the AS. Therefore, there is a total of 24 possible pairs of interaction that could be established upon AS binding.

The binding interaction graph is thus defined by a $24\times 24$ adjacency matrix $A$ where $A_{ij}=A_{ji}=1$, if the two binding interactions $i$ and $j$ are geometrically compatible, and $A_{ij}=0$, otherwise. With this adjacency matrix, the GBS routine is carried out as described in Sec.~\ref{sec:GBS}, and the corresponding MPS-based holographic implementation is carried out as described in Sec.~\ref{sec:network}. The results of the sampling are analyzed to identify the densest sampled subgraph.

\begin{figure}
    \centering
    \begin{subfigure}[b]{\textwidth}
        \centering
        \includegraphics[scale=0.55]{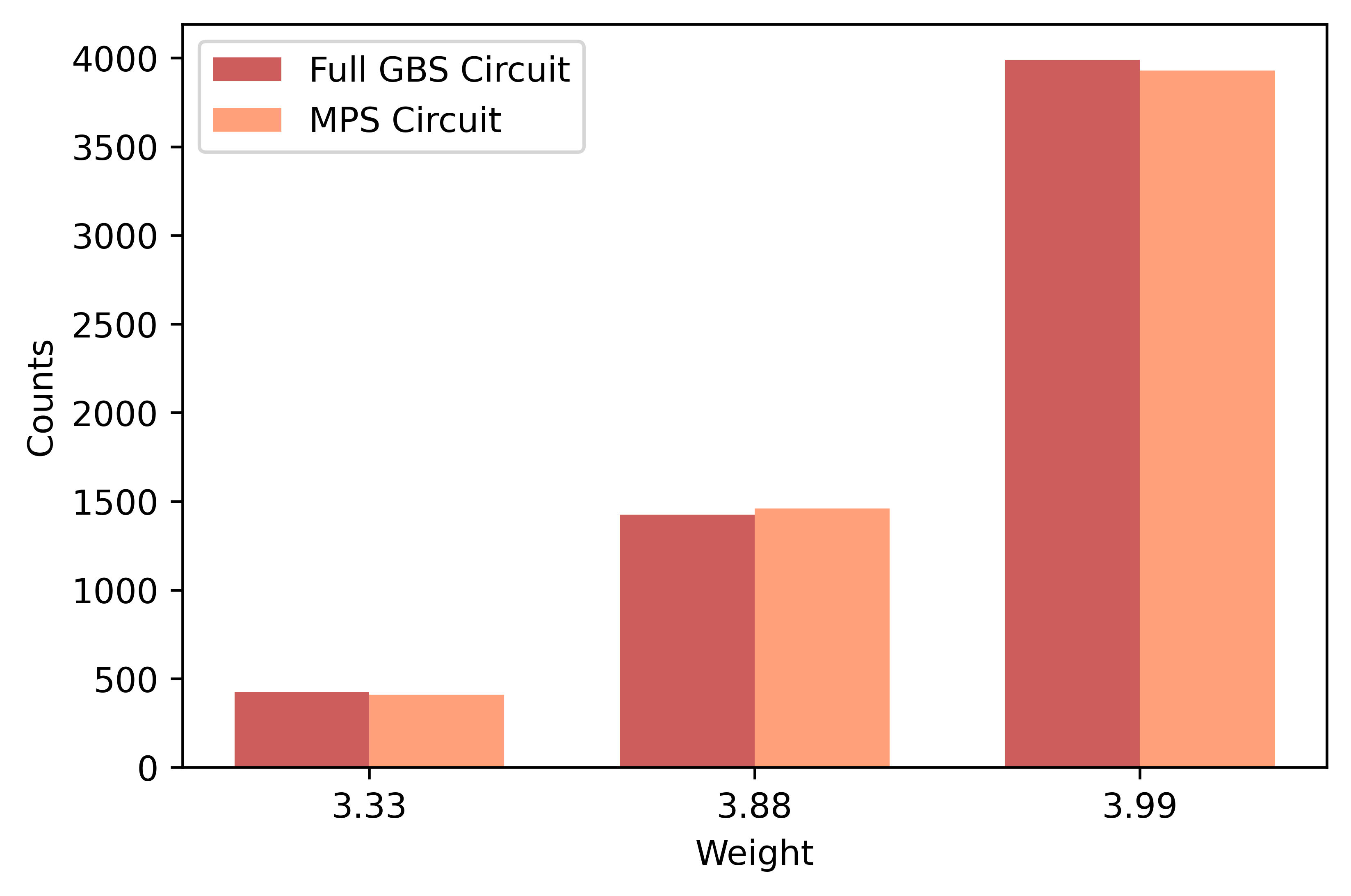}
        \caption{}
    \end{subfigure}
    \begin{subfigure}[b]{\textwidth}
        \centering
        \includegraphics[scale=0.30]{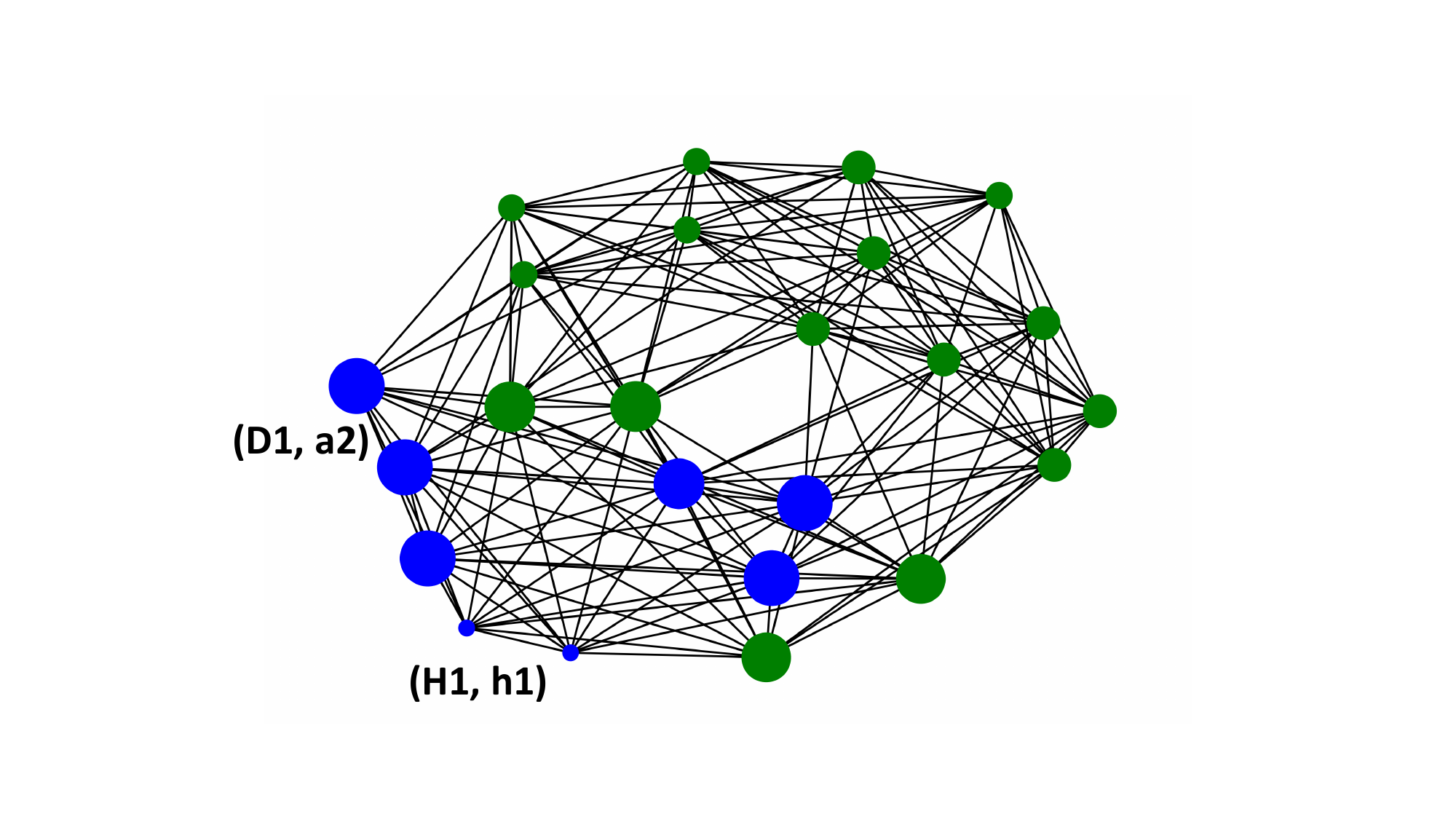}
        \caption{}
    \end{subfigure}
    \begin{subfigure}[b]{\textwidth}
        \centering
        \includegraphics[scale=0.35]{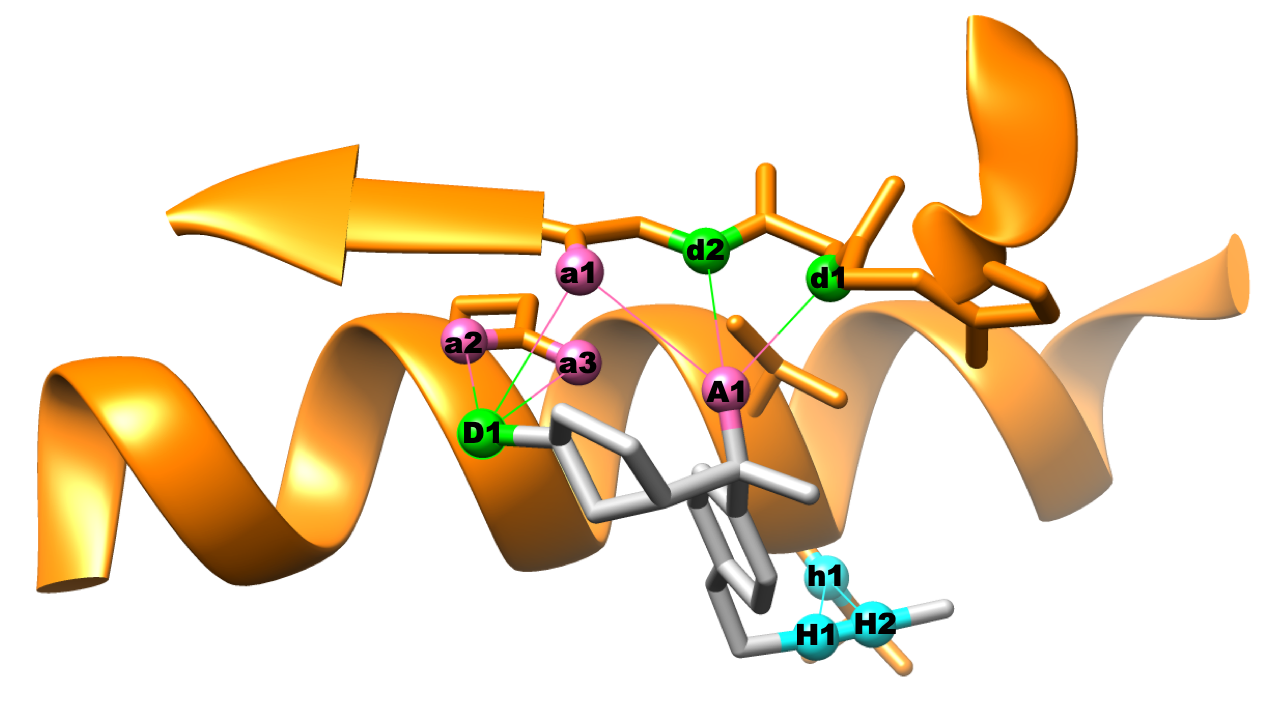}
        \caption{}
    \end{subfigure}
    \caption{(a) Weight of sampled clique from both MPS-based network and full GBS network after postselection. Only three weights (3.33, 3.88 and 3.99) are selected with high frequency. (b) Subgraph for the heaviest (weight=3.99) clique obtained in (a). (c) Visualization of binding interactions of the subgraph in (b).}
    \label{fig:TACE_post}
\end{figure}

Figure~\ref{fig:TACE_post} shows the result of GBS versus MPS-network sampling. Dense subgraphs sampled by GBS are converted to heaviest cliques representing the most probable binding patterns, using the algorithm for post-processing GBS data described in Ref.~\citenum{Banchi2020}. The algorithm first shrinks the GBS sampled subgraph into smaller cliques by sequentially removing vertices with small degree, until finding a clique. Then, the found clique is locally expanded into a large clique according to a local search algorithm that expand the clique by one vertex that is fully connected with all vertices in the original clique. As shown in Fig.~\ref{fig:TACE_post}(a), the sampling routines successfully identify the heaviest clique (weight=3.99) with high probability. Moreover, sampling results from the MPS network agree closely with the GBS result, confirming the accuracy of the proposed MPS-based holographic approach. Fig.~\ref{fig:TACE_post}(b) plots the maximum clique as a subgraph in the binding interaction graph, while Fig.~\ref{fig:TACE_post}(c) visualizes the binding interactions that correspond to the maximum clique.
%

\subsection{Scalability analysis with random graphs}\label{sec:rad}
Here, we examine the scalability and accuracy of the MPS-based algorithm described in Sec.~\ref{sec:network} for a series of random graphs, specified by randomly generated adjacency matrices $A$ that satisfy the conditions described at the beginning of Sec.~\ref{sec:GBS}.

First, we examine the capabilities of our variational approach to approximate the full optical network by comparing the covariance matrix at the output of the optimized MPS-based holographic network to the reference covariance matrix generated by the conventional GBS setup (Table~\ref{table:opt-ref}).

The results displayed in Table~\ref{table:opt-ref} indicate that MPS-circuits equipped with either one or two layers of beam-splitters achieve high precision in approximating the GBS output states, showing that the covariance matrix error remains below 5 percent, even for circuits comprising up to 50 modes. 
\begin{table}[htbp]
    \centering
    \caption{Performance of the MPS-based circuits parameterized by the variational covariance matrix optimization. The number of modes for each test is denoted by $\text{M}$. The relative distance between covariance matrices is defined as $||\sigma_{opt}-\sigma_{ref}||/||\sigma_{ref}||$.}\label{table:opt-ref}
    \begin{minipage}{.30\linewidth}
        \centering
        \begin{subtable}{\linewidth}
            \centering
            \caption{1-layer MPS network}
            \begin{tabular}{cc}
                \toprule
                M & \text{Distance}\\
                \midrule
                12 & 0.05\\
                16 & 0.04\\
                20 & 0.04\\
                24 & 0.03\\
                50 & 0.02\\
                \bottomrule
            \end{tabular}
        \end{subtable}
    \end{minipage}
    \begin{minipage}{.30\linewidth}
        \centering
        \begin{subtable}{\linewidth}
            \centering
            \caption{2-layers MPS network}
            \begin{tabular}{cc}
                \toprule
                M & \text{Distance}\\
                \midrule
                12 & 0.04\\
                16 & 0.04\\
                20 & 0.03\\
                24 & 0.02\\
                50 & 0.01\\
                \bottomrule
            \end{tabular}
        \end{subtable}
    \end{minipage}
\end{table}
%

\section{Conclusion}
In this study, we have introduced quantum circuits designed for simulating multi-mode state-vectors on 3D cQED processors, leveraging matrix product state representations. These circuits have been showcased through simulations of molecular docking, specifically focusing on the binding of a thiol-containing aryl sulfonamide ligand to the tumor necrosis factor-$\alpha$ converting enzyme receptor, utilizing holographic Gaussian boson sampling. Our findings reveal the proposed MPS scheme based on cQED devices with only 2 or 3 modes is able to prepare multimode Gaussian states that closely approximate those obtained by the conventional GBS method even for systems with up to 50 modes. This approach opens the door to a broad spectrum of GBS applications that could be efficiently executed on compact 3D cQED processors using the holographic method.

\section{Acknowledgments}
The authors acknowledge support from the NSF grant 2124511 [CCI Phase I: NSF Center for Quantum Dynamics on Modular Quantum Devices (CQD-MQD)].

\section{Code availability}
The python code for the TACE-AS GBS/MPS simulation is available at \href{https://colab.research.google.com/drive/1_cWqkU42e5jCtnwwFfd833eBaYsqhLV6?usp=sharing}{\textcolor{blue}{this link}}.

\appendix
\section{Representations of the Gaussian state}\label{sec:pfunc}
The goal of this section is to first introduce the eigenstates of the annihilation operator, the so-called coherent states $\vert\alpha\rangle$ that fulfill the eigenvalue equation
\begin{align}\label{eq:eigvp}
    \hat{a}\vert\alpha\rangle
    =\alpha\vert\alpha\rangle,
\end{align}
where $\alpha\in\mathbb{C}$ is a given complex number. Based on the coherent states, we will then introduce the so called P-representation of density operators that we will use to unfold GBS. We will see that a coherent state has a precise phase defined by the complex amplitude $\alpha$, although an indefinite number of photons, like the state of coherent light in a laser beam. In contrast, a Fock state is an eigenstate of the number operator, corresponding to a fixed well-defined number of photons although completely arbitrary (random) phase. In the following, we focus on harmonic oscillator coherent states, while noting that generalization to anharmonic coherent states is readily available.\cite{Kais1990}

\subsection{Overview of the coherent states}
The goal of this subsection is to provide an overview of properties of coherent states that would be vital for applications to the GBS.

\paragraph{Displacement operator.}
We start by showing that we can create coherent states, as follows
\begin{align}\label{eq:crecs1}
    \hat{D}(\alpha)\vert 0\rangle
    =\vert\alpha\rangle,
\end{align}
where $\vert 0\rangle$ is the vacuum state defined as the ground state of the harmonic oscillator, and $\hat{D}(\alpha)$ is the displacement operator, defined as follows
\begin{align}\label{eq:displa}
    \begin{split}
        \hat{D}(\alpha)
        &=e^{\alpha\hat{a}^{\dagger}-\alpha^*\hat{a}}\\
        &=e^{\alpha\hat{a}^{\dagger}}e^{-\alpha^*\hat{a}}e^{-\frac{1}{2}\vert\alpha\vert^2}.\\
    \end{split}
\end{align}
The second row of Eq.~(\ref{eq:displa}) is obtained from the first one by using the Hausdorff formula $e^{\hat A+\hat B}=e^{\hat A}e^{\hat B}e^{-\frac{1}{2}[\hat A,\hat B]}$, with $\hat A=\alpha\hat{a}^{\dagger}$ and $\hat B=-\alpha^*\hat{a}$, which is valid if $[\hat A,[\hat A,\hat B]]=[\hat B,[\hat A,\hat B]]=0$, as in this case. Note that $[\hat A,\hat B]=-\vert\alpha\vert^2[\hat{a}^{\dagger},\hat{a}]=\vert\alpha\vert^2$, since $[\hat{a}^{\dagger},\hat{a}]=-1$.

We will obtain Eqs.~(\ref{eq:eigvp}) and~(\ref{eq:crecs1}) by showing that, according to the Baker--Campbell--Hausdorff relation, $\hat{D}(\alpha)^{\dagger}\hat{a}\hat{D}(\alpha)=\hat{a}+\alpha$ (part 1). Hence, with $\hat{a}\vert 0\rangle=0$, we can conclude that $\hat{D}(\alpha)^{\dagger}\hat{a}\hat{D}(\alpha)\vert 0\rangle=\alpha\vert 0\rangle$, and $\hat{a}\hat{D}(\alpha)\vert 0\rangle=\alpha\hat{D}(\alpha)\vert 0\rangle$ (part 2). Consequently, by definition, $\hat{D}(\alpha)\vert 0\rangle$ is equal to $\vert\alpha\rangle$: the eigenfunction of $\hat{a}$ with eigenvalue $\alpha$.\\[2mm]

Let us start with part 1:
Firstly, we show that $\hat{D}(\alpha)^{-1}=\hat{D}(-\alpha)=\hat{D}(\alpha)^{\dagger}$:
\begin{align}\label{eq:inverseD}
    \begin{split}
        \hat{D}(\alpha)^{-1}
        &=e^{\frac{1}{2}\vert\alpha\vert^2}e^{\alpha^*\hat{a}}e^{-\alpha\hat{a}^{\dagger}}\\
        &=e^{-\frac{1}{2}\vert\alpha\vert^2}e^{-\alpha\hat{a}^{\dagger}}e^{\alpha^*\hat{a}}
        =\hat{D}(-\alpha),\\
    \end{split}
\end{align}
where the first equality follows from $\hat{D}(\alpha)^{-1}\hat{D}(\alpha)=1$. The second row of Eq.~(\ref{eq:inverseD}) is obtained from the first one since 
\begin{align}
    e^{\alpha^*\hat{a}}e^{-\alpha\hat{a}^{\dagger}}
    =e^{-\alpha\hat{a}^{\dagger}}e^{\alpha^*\hat{a}}e^{-\vert\alpha\vert^2}.
\end{align}
The Baker--Campbell--Hausdorff relation
\begin{align}
    e^{\hat A}\hat Be^{-\hat A}
    =\hat B+[\hat A,\hat B]+\frac{1}{2}[\hat A,[\hat A,\hat B]]+\dots
\end{align}
can be used with $\hat A=-\alpha\hat{a}^{\dagger}+\alpha^*\hat{a}$, and $\hat B=\hat{a}$ to show that 
\begin{align}\label{eq:Hausdorff}
    \hat{D}(\alpha)^{\dagger}\hat{a}\hat{D}(\alpha)
    =\hat{a}+\alpha,
\end{align}
since $[\hat A,\hat B]=[-\alpha\hat{a}^{\dagger}+\alpha^*\hat{a},\hat{a}]=\alpha$, and therefore $[\hat A,[\hat A,\hat B]]=0$.
For part 2, we apply the vacuum state $\vert 0\rangle$ to Eq.~(\ref{eq:Hausdorff}) and obtain
\begin{align}
    \hat{D}(\alpha)^{\dagger}\hat{a}\hat{D}(\alpha)\vert 0\rangle
    =\alpha\vert 0\rangle,
\end{align}
since we have $\hat{a}\vert 0\rangle=0$.
Therefore, according to Eq.~(\ref{eq:eigvp}), we conclude that
\begin{align}\label{eq:Hausdorff3}
    \hat{D}(\alpha)\vert 0\rangle
    =\vert\alpha\rangle,
\end{align}
which indeed is Eq.~(\ref{eq:crecs1}).

\paragraph{Series expansion.}
Substituting Eq.~(\ref{eq:displa}) into Eq.~(\ref{eq:Hausdorff3}), we obtain
\begin{align}\label{eq:Hausdorff33}
    \vert\alpha\rangle
    =e^{-\frac{1}{2}\vert\alpha\vert^2}e^{\alpha\hat{a}^{\dagger}}e^{-\alpha^*\hat{a}}\vert 0\rangle
\end{align}
and by expanding the exponentials in Taylor series, we get
\begin{align}\label{eq:Ha}
    \begin{split}
        \vert\alpha\rangle
        &=e^{-\frac{1}{2}\vert\alpha\vert^2}e^{\alpha\hat{a}^{\dagger}}\vert 0\rangle\\
        &=e^{-\frac{1}{2}\vert\alpha\vert^2}\sum_{n=0}^{\infty}\frac{\alpha^n}{n!}(\hat{a}^\dagger)^n\vert 0\rangle\\
        &=e^{-\frac{1}{2}\vert\alpha\vert^2}\sum_{n=0}^{\infty}\frac{\alpha^n}{\sqrt{n!}}\vert n\rangle,
    \end{split}
\end{align}
where the third row is obtained by $\hat{a}^{\dagger}\vert n\rangle=\sqrt{n+1}\vert n+1\rangle$.
In particular, we obtain the following eigenvalue equation $\hat{a}\vert\alpha\rangle=\alpha\vert\alpha\rangle$ since
\begin{align}
    \begin{split}
        \hat{a}\vert\alpha\rangle
        &=e^{-\frac{1}{2}\vert\alpha\vert^2}\sum_{n=1}^{\infty}\frac{\alpha^n}{\sqrt{n !}}\sqrt{n}\vert n-1\rangle\\
        &=e^{-\frac{1}{2}\vert\alpha\vert^2}\sum_{n=1}^{\infty}\alpha\frac{\alpha^{n-1}}{\sqrt{(n-1) !}}\vert n-1\rangle\\
        &=\alpha e^{-\frac{1}{2}\vert\alpha\vert^2}\sum_{n=0}^{\infty}\frac{\alpha^n}{\sqrt{(n) !}}\vert n\rangle.
    \end{split}
\end{align}
Given that the coherent state is an eigenstate of the annihilation operator, its complex conjugate is an eigenstate of the creation operator
\begin{align}
    \left(\hat{a}\left|\alpha\right\rangle\right)^{\dagger}
    &=\left(\alpha\left|\alpha\right\rangle\right)^{\dagger},\\
    \left\langle\alpha\right|\hat{a}^{\dagger}
    &=\left\langle\alpha\right|\alpha^{\star}.
\end{align}
%

\paragraph{Overlap.}
Coherent states are not orthogonal since, according to Eq.~(\ref{eq:Ha}),
\begin{align}\label{eq:overlap21}
    \begin{split}
        \langle\beta\vert\alpha\rangle
        &=e^{-\frac{1}{2}\vert\alpha\vert^2}e^{-\frac{1}{2}\vert\beta\vert^2}\sum_{n=0}^{\infty}\sum_{m=0}^{\infty}\frac{(\beta^*)^m\alpha^n}{\sqrt{m!n!}}\langle m\vert n\rangle\\
        &=e^{-\frac{1}{2}\vert\alpha\vert^2}e^{-\frac{1}{2}\vert\beta\vert^2}\sum_{n=0}^{\infty}\frac{(\beta^*)^n\alpha^n}{n!}\\
        &=e^{-\frac{1}{2}\vert\alpha\vert^2}e^{-\frac{1}{2}\vert\beta\vert^2}e^{\beta^*\alpha}\\
        &=e^{-\frac{1}{2}\vert\beta-\alpha\vert^2}e^{\frac{1}{2}(\beta^*\alpha-\beta\alpha^*)} 
    \end{split}
\end{align} 
%

\paragraph{Expectation values.}
The expectation value of position,
\begin{align}
    \hat{x}
    =\sqrt{\frac{\hbar}{2m\omega}}\left(a+a^{\dagger}\right),
\end{align}
follows from
\begin{align}\label{eq:ree}
    \left\langle\alpha\left|\hat{x}\right|\alpha\right\rangle
    &=\left\langle\alpha\left|\sqrt{\frac{\hbar}{2m\omega}}\left(a+a^{\dagger}\right)\right|\alpha\right\rangle\\
    &=\sqrt{\frac{\hbar}{2m\omega}}\left(\left\langle\alpha\left|a\right|\alpha\right\rangle+\left\langle\alpha\left|a^{\dagger}\right|\alpha\right\rangle\right)\\
    &=\sqrt{\frac{\hbar}{2m\omega}}\left(\alpha\left\langle\alpha\middle|\alpha\right\rangle+\alpha^{\star}\left\langle\alpha\middle|\alpha\right\rangle\right)\\
    &=\sqrt{\frac{\hbar}{2m\omega}}\left(\alpha+\alpha^{\star}\right)\\
    &=\sqrt{\frac{2\hbar}{m\omega}}\,\text{Re}\left(\alpha\right).
\end{align}
Likewise, the expectation value of the momentum,
\begin{align}
    \hat{p}
    =-i\sqrt{\frac{m\hbar\omega}{2}}\left(a-a^{\dagger}\right),
\end{align}
follows from
\begin{align}\label{eq:ime}
    \left\langle\alpha\left|\hat{p}\right|\alpha\right\rangle
    &=\left\langle\alpha\left|-i\sqrt{\frac{m\hbar\omega}{2}}\left(a-a^{\dagger}\right)\right|\alpha\right\rangle\\
    &=-i\sqrt{\frac{m\hbar\omega}{2}}\left(\left\langle\alpha\left|a\right|\alpha\right\rangle-\left\langle\alpha\left|a^{\dagger}\right|\alpha\right\rangle\right)\\
    &=-i\sqrt{\frac{m\hbar\omega}{2}}\left(\alpha\left\langle\alpha\middle|\alpha\right\rangle-\alpha^{\star}\left\langle\alpha\middle|\alpha\right\rangle\right)\\
    &=-i\sqrt{\frac{m\hbar\omega}{2}}\left(\alpha-\alpha^{\star}\right)\\
    &=\sqrt{2m\hbar\omega}\,\text{Im}\left(\alpha\right).
\end{align}
Therefore, according to Eq.~(\ref{eq:ree}) and Eq.~(\ref{eq:ime}), we obtain
\begin{align}\label{eq:alf}
    \alpha
    =\alpha_r+i\alpha_i
    =\sqrt{\frac{m\omega}{2\hbar}}q_{\alpha}+i\frac{1}{\sqrt{2m\hbar\omega}}p_{\alpha},
\end{align}
where $q_{\alpha}=\langle\alpha\vert\hat{x}\vert\alpha\rangle$ and $p_{\alpha}=\langle\alpha\vert\hat{p}\vert\alpha\rangle$.

\paragraph{Representation as wavefunction.}
The wavefunctions can be obtain by substituting the eigenfunctions of the Harmonic oscillator,
\begin{align}
    \langle x\vert n\rangle
    =(2^nn!)^{-1/2}\left(\frac{m\omega}{\pi\hbar}\right)^{1/4}\exp(-\tilde{x}^2/2)H_n(\tilde{x}),
\end{align}
into Eq.~(\ref{eq:Ha}), where $\tilde{x}=x\sqrt{m\omega/\hbar}$, with $H_n$ the $n$th-Hermite polynomial, giving
\begin{align}\label{eq:Hausdorff334}
    \langle x\vert\alpha\rangle
    =\left(\frac{m\omega}{\pi\hbar}\right)^{1/4}e^{-\frac{1}{2}\vert\alpha\vert^2}e^{-\frac{1}{2}\tilde{x}^2}\sum_{n=0}^{\infty}\frac{(\alpha/\sqrt{2})^n}{n!}H_n(\tilde{x}).
\end{align}
The vacuum state corresponding to $n=0$ photons is described by the wavefunction $\langle x\vert 0\rangle$. According to Eq.~(\ref{eq:Hausdorff334}), $\langle x\vert 0\rangle$ is also a coherent state with $\alpha=0$ ({\em i.e.}, the ground state of a harmonic oscillator with mass $m$ and frequency $\omega$). We note that the Hermite polynomials are given by the characteristic function
\begin{align}
    e^{2\tilde{x}t-t^2}
    =\sum_{n=0}^{\infty}H_n(\tilde{x})\frac{t^n}{n!},
\end{align}
as can be verified by the Taylor expansion at $\tilde{x}$. Therefore, substituting the characteristic function into Eq.~(\ref{eq:Hausdorff334}), with $t=\alpha/\sqrt{2}$, we obtain 
\begin{align}\label{eq:Hausdorff335}
    \begin{split}
        \langle x\vert\alpha\rangle 
        &=\left(\frac{m\omega}{\pi\hbar}\right)^{1/4}e^{-\frac{1}{2}\vert\alpha\vert^2}e^{\frac{1}{2}\tilde{x}^2}e^{-(\tilde{x}-\alpha/\sqrt{2})^2}\\
        &=\left(\frac{m\omega}{\pi\hbar}\right)^{1/4}e^{-\frac{1}{2}(\alpha_r^2+\alpha_i^2)}e^{\frac{1}{2}\tilde{x}^2}e^{-(\tilde{x}-\alpha_r/\sqrt{2}-i\alpha_i/\sqrt{2})^2}\\
        &=\left(\frac{m\omega}{\pi\hbar}\right)^{1/4}e^{-\frac{1}{2}(\alpha_r^2+\alpha_i^2)}e^{\frac{1}{2}\tilde{x}^2}e^{-(\tilde{x}^2+\alpha_r^2/2-\alpha_i^2/2+i\alpha_r\alpha_i-\sqrt{2}\tilde{x} (\alpha_r+i\alpha_i))}\\
        &=\left(\frac{m\omega}{\pi\hbar}\right)^{1/4}e^{-\alpha_r^2}e^{-\frac{1}{2}\tilde{x}^2}e^{\sqrt{2}\tilde{x}\alpha_r}e^{-i\alpha_i(\alpha_r-\sqrt{2}\tilde{x})}\\
        &=\left(\frac{m\omega}{\pi\hbar}\right)^{1/4}e^{-(\alpha_r-\tilde{x}/\sqrt{2})^2}e^{-i\alpha_i(\alpha_r-\sqrt{2}\tilde{x})}.
    \end{split}
\end{align}
Substituting $\alpha_r$ and $\alpha_i$ in terms of $q_{\alpha}$ and $p_{\alpha}$, we obtain
\begin{align}
    \begin{split}
        \langle x\vert\alpha\rangle
        &=\left(\frac{m\omega}{\pi\hbar}\right)^{1/4}e^{-\left(q_{\alpha}\sqrt{\frac{m\omega}{2\hbar}}-x\sqrt{\frac{m\omega}{2\hbar}}\right)^2}e^{-ip_{\alpha}\sqrt{\frac{1}{2m\hbar\omega}}\left(q_{\alpha}\sqrt{\frac{m\omega}{2\hbar}}-\sqrt{2}x\sqrt{\frac{m\omega}{\hbar}}\right)}\\
        &=\left(\frac{m\omega}{\pi\hbar}\right)^{1/4}e^{-\left(\frac{m\omega}{2\hbar}\right)\left(x-q_{\alpha}\right)^2}e^{\frac{i}{\hbar}p_{\alpha}(x-q_{\alpha})}e^{\frac{i}{2\hbar}p_{\alpha} q_{\alpha}}\\
        &=\left(\frac{\gamma}{\pi}\right)^{1/4}e^{-\frac{\gamma}{2}\left(x-q_{\alpha}\right)^2}e^{\frac{i}{\hbar}p_{\alpha}(x-q_{\alpha})}e^{\frac{i}{2\hbar}p_{\alpha}q_{\alpha}},
    \end{split}
\end{align}
with $\gamma=m\omega/\hbar$.

\subsection{P-representation of the density operator}\label{sec:pfunc2}
The \color{blue}\href{https://en.wikipedia.org/wiki/Roy_J._Glauber}{Glauber}-\href{https://en.wikipedia.org/wiki/E._C._George_Sudarshan}{Sudarshan} \color{black} $P$-representation of the density operator, $\hat{\rho}$ ({\em i.e.}, the $P$-function, $P(\alpha)$), is a pseudo-probability distribution defined, as follows
\begin{align}\label{eq:diag}
    \hat{\rho}
    =\int P(\alpha)\vert\alpha\rangle\langle\alpha\vert\,\mathrm{d}^2\alpha,
\end{align}
where $\mathrm{d}^2\alpha=\mathrm{d}\operatorname{Re}(\alpha)\mathrm{d}\operatorname{Im}(\alpha)$.
As shown below,
\begin{align}\label{eq:pcohe0}
    P(\alpha)
    =\frac{e^{\vert\alpha\vert^2}}{\pi^2}\int e^{-\alpha^*u+u^*\alpha}\langle-u\vert\hat{\rho}\vert u\rangle e^{\vert u\vert^2}\,\mathrm{d}^2u.
\end{align}
To prove Eq.~(\ref{eq:pcohe0}), we follow \color{blue}\href{http://ursula.chem.yale.edu/~batista/classes/CHEM584/Mehta67}{Mehta} \color{black} and compute $\langle-u\vert\hat{\rho}\vert u\rangle$ by using Eq.~(\ref{eq:diag}),
\begin{align}
    \langle-u\vert\hat{\rho}\vert u\rangle
    =\int P(\alpha)\langle-u\vert\alpha\rangle\langle\alpha\vert u\rangle\mathrm{d}^2\alpha.
\end{align}
Substituting $\langle\alpha\vert u\rangle$ according to Eq.~(\ref{eq:overlap21}), we obtain
\begin{align}
    \begin{split}
        \langle-u\vert\hat{\rho}\vert u\rangle
        &=\int P(\alpha)e^{-\frac{1}{2}\vert u\vert^2-\frac{1}{2}\vert\alpha\vert^2-u^*\alpha}e^{-\frac{1}{2}\vert u\vert^2-\frac{1}{2}\vert\alpha\vert^2+\alpha^*u}\,\mathrm{d}^2\alpha\\
        &=e^{-\vert u\vert^2}\int P(\alpha)e^{-\vert\alpha\vert^2}e^{\alpha^*u-u^*\alpha}\,\mathrm{d}^2\alpha.
    \end{split}
\end{align}
Introducing the variable substitution $\alpha=x+iy$ and $u=x'+iy'$, we obtain
\begin{align}\label{eq:prei}
    \begin{split}
        \langle-u(x',y')\vert\hat{\rho}\vert u(x',y')\rangle e^{\vert u(x',y')\vert^2}
        &=\int P(\alpha(x,y))e^{-x^2-y^2}e^{(x-iy)(x'+iy')-(x'-iy')(x+iy)}\,\mathrm{d}x\mathrm{d}y\\
        &=\int P(\alpha(x,y))e^{-x^2-y^2}e^{-i2yx'+i2y'x}\,\mathrm{d}x\mathrm{d}y.
    \end{split}
\end{align}
Introducing the function $I(\tilde{x},\tilde{y})$ by
\begin{align}
    \begin{split}
        I(\tilde{x},\tilde{y})
        &=\frac{1}{\pi^2}\int e^{i2x'\tilde{y}-i2y'\tilde{x}}\langle-u(x',y')\vert\hat{\rho}\vert u(x',y')\rangle e^{\vert u(x',y')\vert^2}\,\mathrm{d}x'\mathrm{d}y'\\
        &=\frac{1}{\pi^2}\int P(\alpha(x,y))e^{-x^2-y^2}\int e^{-i2(y-\tilde{y})x'+iy'2 (x-\tilde{x})}\,\mathrm{d}x'\mathrm{d}y'\,\mathrm{d}x\mathrm{d}y\\
        &=\frac{1}{(2\pi)^2}\int P(\alpha(x,y))e^{-x^2-y^2}\int e^{-i(y-\tilde{y})x''+iy''(x-\tilde{x})}\,\mathrm{d}x''\mathrm{d}y''\,\mathrm{d}x\mathrm{d}y\\
        &=\int P(\alpha(x,y))e^{-x^2-y^2}\delta(y-\tilde{y})\delta(x-\tilde{x})\,\mathrm{d}x\mathrm{d}y,
    \end{split}
\end{align}
we conclude that
\begin{align}
    I(\tilde{x},\tilde{y})
    =P(\alpha(\tilde{x},\tilde{y}))e^{-\tilde{x}^2-\tilde{y}^2},
\end{align}
which gives us
\begin{align}\label{eq:pcohe}
    \begin{split}
        P(\alpha({x},{y}))
        &=\frac{e^{x^2+y^2}}{\pi^2}\int e^{ix'{y}-iy'{x}}\langle-u(x',y')\vert\hat{\rho}\vert u(x',y')\rangle e^{\vert u(x',y')\vert^2}\,\mathrm{d}x'\mathrm{d}y',\\
        P(\alpha)
        &=\frac{e^{\vert\alpha\vert^2}}{\pi^2}\int e^{-\alpha^*u+u^*\alpha}\langle-u\vert\hat{\rho}\vert u\rangle e^{\vert u\vert^2}\,\mathrm{d}^2u.
    \end{split}
\end{align}
%

\paragraph{Pure coherent states.}
For the pure state $\hat{\rho}=\vert\beta\rangle\langle\beta\vert$, we obtain according to Eq.~(\ref{eq:overlap21})
\begin{align}\label{eq:muu}
    \begin{split}
        \langle-u\vert\hat{\rho}\vert u\rangle
        &=e^{-\frac{1}{2}\vert-u-\beta\vert^2}e^{\frac{1}{2}(-u^*\beta+u\beta^*)}e^{-\frac{1}{2}\vert\beta-u\vert^2}e^{\frac{1}{2}(\beta^*u-\beta u^*)}\\
        &=e^{-\vert u\vert^2}e^{-\vert\beta\vert^2}e^{u\beta^*-u^*\beta}.
    \end{split}
\end{align}
Therefore, substituting Eq.~(\ref{eq:muu}) into Eq.~(\ref{eq:pcohe}), we obtain
\begin{align}
    \begin{split}
        P(\alpha)
        &=\frac{e^{\vert\alpha\vert^2}}{\pi^2}\int e^{-\alpha^*u+u^*\alpha}e^{-\vert u\vert^2}e^{-\vert\beta\vert^2}e^{u\beta^*-u^*\beta}e^{\vert u\vert^2}\,\mathrm{d}^2u\\
        &=\frac{e^{\vert\alpha\vert^2}e^{-\vert\beta\vert^2}}{\pi^2}\int e^{-\alpha^*u+u^*\alpha}e^{u\beta^*-u^*\beta}\,\mathrm{d}^2u\\
        &=\frac{e^{\vert\alpha\vert^2}e^{-\vert\beta\vert^2}}{\pi^2}\int e^{-u(\alpha-\beta)^*+u^*(\alpha-\beta)}\,\mathrm{d}^2u,
    \end{split}
\end{align}
and considering that $u=x'+iy'$, we obtain
\begin{align}\label{eq:deltas}
    \begin{split}
        P(\alpha)
        &=\frac{e^{\vert\alpha\vert^2}e^{-\vert\beta\vert^2}}{\pi^2}\int e^{i2x'\operatorname{Im}(\alpha-\beta)-i2y'\operatorname{Re}(\alpha-\beta)}\,\mathrm{d}x'\mathrm{d}y'\\
        &=\frac{e^{\vert\alpha\vert^2}e^{-\vert\beta\vert^2}}{(2\pi)^2}\int e^{ix''\operatorname{Im}(\alpha-\beta)-iy''\operatorname{Re}(\alpha-\beta)}\,\mathrm{d}x''\mathrm{d}y''\\
        &=e^{\vert\alpha\vert^2}e^{-\vert\beta\vert^2}\delta(\operatorname{Im}(\alpha-\beta))\delta(\operatorname{Re}(\alpha-\beta))\\
        &=\delta^2(\alpha-\beta).
    \end{split}
\end{align}
Hence, for a pure coherent state, $P(\alpha)$ coincides with the classical density of states. In particular, this shows that coherent states are classical-like quantum states.

\paragraph{Pure number states.}
The P-representation of a pure number state, $\hat{\rho}=\vert n\rangle\langle n\vert$, is obtained, as follows
\begin{align}\label{eq:muum}
    \begin{split}
        \langle-u\vert\hat{\rho}\vert u\rangle
        &=\langle-u\vert n\rangle\langle n\vert u\rangle\\
        &=e^{-\vert u\vert^2}\frac{u^n}{n!}(-u^*)^n,
    \end{split}
\end{align}
where we have substituted $\langle n\vert u\rangle$ in the second row, according to Eq.~(\ref{eq:Ha}), as follows
\begin{align}\label{eq:Has}
    \langle n\vert u\rangle 
    =e^{-\frac{1}{2}\vert u\vert^2}\frac{u^n}{\sqrt{n!}}.
\end{align}
Therefore, substituting Eq.~(\ref{eq:muum}) into Eq.~(\ref{eq:pcohe}), we obtain
\begin{align}\label{eq:pcohe2}
    \begin{split}
        P(\alpha)
        &=\frac{e^{\vert\alpha\vert^2}}{\pi^2}\int e^{-\alpha^*u+u^*\alpha}\langle-u\vert\hat{\rho}\vert u\rangle e^{\vert u\vert^2}\,\mathrm{d}^2u\\
        &=\frac{e^{\vert\alpha\vert^2}}{\pi^2}\int e^{-\alpha^*u+u^*\alpha}e^{-\vert u\vert^2}\frac{(-uu^*)^n}{n !}e^{\vert u\vert^2}\,\mathrm{d}^2u\\
        &=\frac{e^{\vert\alpha\vert^2}}{n!\pi^2}\int e^{-\alpha^*u+u^*\alpha}(-uu^*)^n\,\mathrm{d}^2u\\
        &=\frac{e^{\vert\alpha\vert^2}}{n!\pi^2}\frac{\partial^{2n}}{\partial^{n}\alpha\partial^{n}\alpha^*}\int e^{-\alpha^*u+u^*\alpha}\,\mathrm{d}^2u,
    \end{split}
\end{align}
which can also be written as
\begin{align}\label{eq:pcohe3}
    P(\alpha)
    =\frac{e^{\vert\alpha\vert^2}}{n!}\frac{\partial^{2n}}{\partial^{n}\alpha\partial^{n}\alpha^*}\delta^2{\alpha}.
\end{align}
Analogously, for an $M$-mode Gaussian state, we obtain
\begin{align}\label{eq:pfunc2}
    P_{\bigotimes_{j=1}^M{\vert n_j\rangle\langle n_j\vert}}(\alpha)
    =\prod_{j=1}^M\frac{e^{\vert\alpha_j\vert^2}}{n_j!}\frac{\partial^{2n_j}}{\partial^{n_j}\alpha_j\partial^{n_j}\alpha_j^*}\delta^2{\alpha_j}.
\end{align}
We note that this is the so-called {\em tempered distribution} function, which operates only as the argument of an integral, as follows
\begin{align}\label{eq:pcoh33}
    \int F(\alpha)\frac{\partial^{2n}}{\partial^{n}\alpha\partial^{n}\alpha^*}\delta^2{\alpha}\,\mathrm{d}^2{\alpha}
    =\frac{\partial^{2n}F(\alpha)}{\partial\alpha^n\partial^n\alpha^*}\Bigg|_{\alpha=0,\,\alpha^*=0}.
\end{align}
%

\paragraph{Expectation values for operators.}
In general, the $P$-representation of an operator $\hat{O}(\hat{a}^{\dagger},\hat{a})$, is analogous to the representation of the density operator introduced by Eq.~(\ref{eq:diag}). It involves the $P$-function $P_{\hat{O}}(\alpha)$, which is defined, as follows
\begin{align}
    \hat{O}
    =\int P_{\hat{O}}(\alpha)\vert\alpha\rangle\langle\alpha\vert\,\mathrm{d}^2\alpha,
\end{align}
with expectation value given by
\begin{align}
    \begin{split}
        \langle\hat{O}\rangle
        &=\text{Tr}[\hat{O}\hat{\rho}]\\
        &=\sum_n\int P_{\hat{O}}(\alpha)\langle n\vert\alpha\rangle\langle\alpha\vert\hat{\rho}\vert n\rangle\,\mathrm{d}^2\alpha\\
        &=\int P_{\hat{O}}(\alpha)\sum_n\langle\alpha\vert\hat{\rho}\vert n\rangle\langle n\vert\alpha\rangle\,\mathrm{d}^2\alpha\\
        &=\int P_{\hat{O}}(\alpha)\langle\alpha\vert\hat{\rho}\vert\alpha\rangle\,\mathrm{d}^2\alpha\\
        &=\pi\int P_{\hat{O}}(\alpha)Q(\alpha)\,\mathrm{d}^2\alpha,
    \end{split}
\end{align}
where $Q(\alpha)=\pi^{-1}\langle\alpha\vert\hat{\rho}\vert\alpha\rangle$ is called the Husimi function. In particular, for a pure state $\hat{\rho}=\vert\psi\rangle\langle\psi\vert$, the Husimi function is $Q(\alpha)=\pi^{-1}\vert\langle\psi\vert\alpha\rangle\vert^2$. The derivation of the Husimi function for an multimode Gaussian will be the subject of the following subsection.
For the particular case $\hat{O}=\vert n\rangle\langle n\vert$, we obtain:
\begin{align}\label{eq:expv}
    \text{Tr}[\hat{O}\hat{\rho}]
    =\text{Tr}[\hat{\rho}\vert n\rangle\langle n\vert]
    =\pi\int P_{{\vert n\rangle\langle n\vert}}(\alpha)Q(\alpha)\,d^2\alpha,
\end{align}
where $P_{{\vert n\rangle\langle n\vert}}(\alpha)$ is defined according to Eq.~(\ref{eq:pcohe3}) as
\begin{align}\label{eq:pfunc}
    P_{{\vert n\rangle\langle n\vert}}(\alpha)
    =\frac{e^{\vert\alpha\vert^2}}{n!}\frac{\partial^{2n}}{\partial^{n}\alpha\partial^{n}\alpha^*}\delta^2{\alpha},
\end{align}
which yields 
\begin{align}\label{eq:pcc}
    \text{Tr}[\hat{\rho}\vert n\rangle\langle n\vert]
    =\pi\int Q(\alpha)\frac{e^{\vert\alpha\vert^2}}{n!}\frac{\partial^{2n}}{\partial^{n}\alpha\partial^{n}\alpha^*}\delta^2{\alpha}\,\mathrm{d}^2{\alpha}.
\end{align}
%

\subsection{Husimi function of an $M$-mode Gaussian state}
The Husimi function of an $M$-mode Gaussian state is defined, as follows 
\begin{align}\label{eq:qfunc}
    Q(\alpha)
    =\pi^{-M}\langle\alpha\vert\hat{\rho}\vert\alpha\rangle,
\end{align}
where $\alpha=(\alpha_1,\dots,\alpha_M,\alpha_1^*,\dots,\alpha_M^*)^T$. The goal of this subsection is to introduce the Wigner transform for the evaluation of Eq.~\eqref{eq:qfunc} for $M=1$.
As a result of this, we will obtain the following expression
\begin{align}\label{eq:qfunc2}
    Q(\alpha) 
    =\frac{\pi^{-1}}{\sqrt{|\det(\sigma_Q)|}}e^{-\alpha^\dagger\sigma_Q^{-1}\alpha},
\end{align}
where $\sigma_Q=\sigma+I_2/2$. 

\paragraph{Wigner transform.}
For the derivation of Eq.~\eqref{eq:qfunc2}, we first give the Husimi function for a single-mode pure Gaussian state. The elements of the density operator of a single-mode pure Gaussian state are given by
\begin{align}
    \begin{split}
        \langle x\vert\hat{\rho}\vert x'\rangle
        &=\langle x\vert\psi\rangle\langle\psi\vert x'\rangle\\
        &=\left(\frac{\gamma}{\pi}\right)^{1/2}e^{-\frac{\gamma}{2}((x-d_x)^2+(x'-d_x)^2)+\frac{i}{\hbar}d_p(x-x')},
    \end{split}
\end{align}
where $d_x=\langle\hat{x}\rangle$, $d_p=\langle\hat{p}\rangle$, and $\gamma=(\langle(\hat{x}-d_x)^2\rangle)^{-1}/2$. These elements can be Wigner transformed, as follows
\begin{align}
    \begin{split}
        \rho_W(x,p)
        &=(2\pi\hbar)^{-1}\int e^{\frac{i}{\hbar}py}\left\langle x-\frac{y}{2}\vert\hat{\rho}\vert x+\frac{y}{2}\right\rangle\,\mathrm{d}y\\
        &=(2\pi\hbar)^{-1}\int\left(\frac{\gamma}{\pi}\right)^{1/2}e^{-\frac{\gamma}{2}((x-\frac{y}{2}-d_x)^2+(x+\frac{y}{2}-d_x)^2)+\frac{i}{\hbar}(p-d_p)y}\,\mathrm{d}y\\
        &=(2\pi\hbar)^{-1}e^{-\gamma(x^2-2xd_x+d_x^2)}\int\left(\frac{\gamma}{\pi}\right)^{1/2}e^{-\frac{\gamma}{4}y^2+\frac{i}{\hbar}(p-d_p)y}\,\mathrm{d}y\\
        &=(\pi\hbar)^{-1}e^{-\left(\gamma(x-d_x)^2+(p-d_p)^2/(\hbar^2\gamma)\right)}.
    \end{split}
\end{align}
For a given positive constant $\gamma_\alpha>0$, let $z=\left(\sqrt{\gamma_\alpha}(x-d_x),(p-d_p)/(\hbar\sqrt{\gamma_\alpha})\right)^T$, and 
\begin{align}
    \sigma
    =
    \begin{bmatrix}
        \gamma_\alpha/\gamma & 0\\
        0 & \gamma/\gamma_\alpha
    \end{bmatrix}.
\end{align}
We then obtain
\begin{align}\label{eq:wtra}
    \begin{split}
        \rho_W(x,p) 
        &=\frac{e^{-z^T\sigma^{-1}z}}{\pi\hbar}\\
        &=\frac{e^{-\frac{1}{2}z^T\tilde{\sigma}^{-1}z}}{2\pi\hbar\sqrt{|\det(\tilde{\sigma}|)}},
    \end{split}
\end{align}
where $\tilde{\sigma}^{-1}=2\sigma^{-1}$.

\paragraph{Expectation values.}
Expectation values can be calculated in terms of the Wigner transform, as follows
\begin{align}
    \begin{split}
        \langle\hat{O}\rangle
        &=\text{Tr}[\hat{\rho}\hat{O}]\\
        &=\int\langle\tilde{x}\vert\hat{\rho}\vert\tilde{x}'\rangle\langle\tilde{x}'\vert\hat{O}\vert\tilde{x}\rangle\,\mathrm{d}\tilde{x}\mathrm{d}\tilde{x}'\\
        &=\int\left\langle x-\frac{y}{2}\vert\hat{\rho}\vert x+\frac{y}{2}\right\rangle\left\langle x+\frac{y}{2}\vert\hat{O}\vert x-\frac{y}{2}\right\rangle\,\mathrm{d}x\mathrm{d}y,
    \end{split}
\end{align}
where $\tilde{x}=x-\frac{y}{2}$ and $\tilde{x}'=x+\frac{y}{2}$. Therefore,
\begin{align}\label{eq:285}
    \begin{split}
        \langle\hat{O}\rangle
        &=\int\left\langle x-\frac{y}{2}\vert\hat{\rho}\vert x+\frac{y}{2}\right\rangle\int\left\langle x+\frac{y'}{2}\vert\hat{O}\vert x-\frac{y'}{2}\right\rangle\delta(y-y')\,\mathrm{d}y'\,\mathrm{d}x\mathrm{d}y\\
        &=\frac{1}{2\pi\hbar}\int\int\left\langle x-\frac{y}{2}\vert\hat{\rho}\vert x+\frac{y}{2}\right\rangle e^{\frac{i}{\hbar}py}\int\left\langle x+\frac{y'}{2}\vert\hat{O}\vert x-\frac{y'}{2}\right\rangle e^{-\frac{i}{\hbar}py'}\,\mathrm{d}y'\,\mathrm{d}y\,\mathrm{d}x\mathrm{d}p\\
        &=\int\rho_W(x,p)\int\left\langle x+\frac{y'}{2}\vert\hat{O}\vert x-\frac{y'}{2}\right\rangle e^{-\frac{i}{\hbar}py'}\,\mathrm{d}y'\,\mathrm{d}x\mathrm{d}p\\
        &=\int\rho_W(x,p)\int\left\langle x-\frac{y'}{2}\vert\hat{O}\vert x+\frac{y'}{2}\right\rangle e^{\frac{i}{\hbar}py'}\,\mathrm{d}y'\,\mathrm{d}x\mathrm{d}p\\
        &=\int\rho_W(x,p)O_W(x,p)\,\mathrm{d}x\mathrm{d}p,
    \end{split}
\end{align}
where $O_W(x,p)=\int\left\langle x-\frac{y'}{2}\vert\hat{O}\vert x+\frac{y'}{2}\right\rangle e^{\frac{i}{\hbar}py'}\,\mathrm{d}y'$. In particular, the expectation value $\langle\alpha\vert\hat{\rho}\vert\alpha\rangle$ is given by
\begin{align}\label{eq:rhoal}
    \langle\alpha\vert\hat{\rho}\vert\alpha\rangle
    =\int\rho_W(x,p)W_{\alpha}(x,p)\,\mathrm{d}x\mathrm{d}p,
\end{align}
with $\vert\alpha\rangle$ defined according to Eq.~(\ref{eq:Hausdorff335}), and
\begin{align}\label{eq:wal}
    \begin{split}
        W_{\alpha}(x,p)
        &=\int\left\langle x-\frac{y'}{2}\vert\alpha\right\rangle\left\langle\alpha\vert x+\frac{y'}{2}\right\rangle e^{\frac{i}{\hbar}py'}\,\mathrm{d}y'\\
        &=2e^{-\gamma_{\alpha}(x-q_{\alpha})^2}e^{-(p-p_{\alpha})^2/(\gamma_{\alpha}\hbar^2)},
    \end{split}
\end{align}
with $\gamma_{\alpha}=m\omega/\hbar$, $q_{\alpha}=\langle\alpha\vert\hat{x}\vert\alpha\rangle$, and $p_{\alpha}=\langle\alpha\vert\hat{p}\vert\alpha\rangle$, corresponding to the value $\alpha=\sqrt{\gamma_{\alpha}/2}q_{\alpha}+i(2\gamma_{\alpha})^{-1/2}\hbar^{-1}p_{\alpha}$, according to Eq.~(\ref{eq:alf}). Analogously, we introduce $\beta=\sqrt{\gamma_{\alpha}/2}x+i(2\gamma_{\alpha})^{-1/2}\hbar^{-1}p$, from which we conclude that
\begin{align}
    2\vert\beta-\alpha\vert^2
    =\gamma_{\alpha}(x-q_{\alpha})^2+(p-p_{\alpha})^2/(\gamma_{\alpha}\hbar^2),
\end{align}
which shows that
\begin{align}\label{eq:wal2}
    W_{\alpha}(x,p)
    =2e^{-2\vert\beta-\alpha\vert^2}.
\end{align}
Substituting Eqs.~(\ref{eq:wtra}) and (\ref{eq:wal2}) into Eq.~(\ref{eq:rhoal}), we find:
\begin{align}\label{eq:rra0}
    \begin{split}
        \langle\alpha\vert\hat{\rho}\vert\alpha\rangle
        &=\int\frac{e^{-\frac{1}{2}z^T\tilde{\sigma}^{-1}z}}{\pi\hbar\sqrt{|\det(\tilde{\sigma})|}}e^{-2\vert\beta-\alpha\vert^2}\,\mathrm{d}x\mathrm{d}p\\
        &=\frac{1}{\pi\sqrt{|\det(\tilde{\sigma})|}}\int e^{-\frac{1}{2}{\beta}^\dagger\tilde{\sigma}^{-1}{\beta}}e^{-2\vert\beta-\alpha\vert^2}\,\mathrm{d}^2\beta.
    \end{split}
\end{align}
For simplicity, let us consider the case where $d_x=d_p=0$.
Using that $\vert\beta-\alpha\vert^2=(\beta-\alpha)^\dagger(\beta-\alpha)=\beta^\dagger\beta-\beta^\dagger\alpha-\alpha^\dagger\beta+\alpha^\dagger\alpha=\beta^\dagger\beta-2\alpha^\dagger\beta+\alpha^\dagger\alpha$, we obtain
\begin{align}\label{eq:rra}
    \begin{split}
        \langle\alpha\vert\hat{\rho}\vert\alpha\rangle
        &=\frac{e^{-2\alpha^{\dagger}\alpha}}{\pi\sqrt{|\det(\tilde{\sigma})|}}\int e^{-\frac{1}{2}\beta^{\dagger}(\tilde{\sigma}^{-1}+4)\beta}e^{4\alpha^{\dagger}\beta}\,\mathrm{d}^2{\beta}\\
        &=\frac{2e^{-2\alpha^{\dagger}\alpha}}{\sqrt{|\det(\tilde{\sigma}^{-1}+4)||\det(\tilde{\sigma})|}}e^{\frac{1}{2} 4\alpha^{\dagger}(\tilde{\sigma}^{-1}+4)^{-1}4\alpha}\\
        &=\frac{2e^{-\alpha^{\dagger}(2-8(\tilde{\sigma}^{-1}+4)^{-1})\alpha}}{\sqrt{|\det(\tilde{\sigma}^{-1}+4)||\det(\tilde{\sigma})|}}.
    \end{split}
\end{align}
Moreover, a short calculation shows that $2-8(\tilde{\sigma}^{-1}+4)^{-1}=(\sigma+I_2/2)^{-1}$, as well as $\sqrt{|\det(\tilde{\sigma}^{-1}+4)||\det(\tilde{\sigma})|}=2\sqrt{|\det(I_2/2+{\sigma})|}$. Therefore, we finally conclude that
\begin{align}\label{eq:rras}
    \begin{split}
        Q(\alpha)
        &=\pi^{-1}\langle\alpha\vert\hat{\rho}\vert\alpha\rangle\\
        &=\pi^{-1}\frac{e^{-\alpha^{\dagger}(\sigma+I_2/2)^{-1}\alpha}}{\sqrt{|\det(\sigma+I_2/2)|}},
    \end{split}
\end{align}
which, when generalized to an $M$-mode Gaussian, gives Eq.~\eqref{eq:qfunc2}.

\section{Output Probabilities of an $M$-mode Gaussian}\label{sec:output}
In this section, we obtain the output probability distribution
\begin{align}
    \text{Pr}(\bar{n})
    =\text{Tr}\left[\hat{\rho}{\bigotimes_{j=1}^M\vert n_j\rangle\langle n_j\vert}\right]
\end{align}
for an $M$-mode input Gaussian, corresponding to occupation numbers of output modes $\bar{n}$ as described by the tensor product of number state operators $\hat{\bar{n}}=\bigotimes_{j=1}^M\hat{n}_j$, where $\hat{n}_j=\vert n_j\rangle\langle n_j\vert$ measures the probability of observing $n_j$ photons in output mode $j$.

Substituting Eq.~(\ref{eq:qfunc2}) and Eq.~(\ref{eq:pfunc}) into Eq.~(\ref{eq:expv}), we obtain
\begin{align}\label{eq:expv2}
    \text{Pr}(\bar{n})
    =\int\frac{1}{\sqrt{|\det(\sigma_Q)|}}e^{-\frac{1}{2}\alpha^{\dagger}\sigma_Q^{-1}\alpha+\alpha^{\dagger}\alpha}\prod_{j=1}^{M}\frac{1}{n_j!}\frac{\partial^{2n_j}}{\partial^{n_j}\alpha_j\partial^{n_j}\alpha_j^*}\delta^2{\alpha_j},
\end{align}
and integration by parts yields
\begin{align}\label{eq:expv3}
    \text{Pr}(\bar{n})
    =\frac{1}{\sqrt{|\det(\sigma_Q)|}}\prod_{j=1}^M\frac{1}{n_j!}\frac{\partial^{2n_j}}{\partial^{n_j}\alpha_j\partial^{n_j}\alpha_j^*}e^{\frac{1}{2}\alpha^{\dagger}(I_{2M}-\sigma_Q^{-1})\alpha}\Bigg|_{\alpha_j=0}.
\end{align}
We note that 
\begin{align}
    \alpha^{\dagger}(I_{2M}-\sigma_Q^{-1})\alpha
    =\alpha^T
    \begin{bmatrix}
        0 & I_M\\
        I_M & 0
    \end{bmatrix}
    (I_{2M}-\sigma_Q^{-1})\alpha,
\end{align}
since is defined as $\alpha=(\alpha_1,\dots,\alpha_M,\alpha_1^*,\dots,\alpha_M^*)^T$. Hence,
\begin{align}\label{eq:expv4}
    \text{Pr}(\bar{n})
    =\frac{1}{\sqrt{|\det(\sigma_Q)|}}\prod_{j=1}^M\frac{1}{n_j!}\frac{\partial^{2n_j}}{\partial^{n_j}\alpha_j\partial^{n_j}\alpha_j^*}e^{\frac{1}{2}\alpha^{T}K\alpha}\Bigg|_{\alpha_j=0},
\end{align}
where we used
\begin{align}\label{eq:adjj}
    K
    =
    \begin{bmatrix}
        0 & I_M\\
        I_M & 0
    \end{bmatrix}
    (I_{2M}-\sigma_Q^{-1}),
\end{align}
as previously defined. In particular, for the specific case of measuring $n_j\in\{0,1\}$ photons at each output mode, with $N=\sum_{j=1}^M n_j$, we obtain
\begin{align}\label{eq:expv5}
    \begin{split}
        \text{Pr}(\bar{n})
        &=\frac{1}{\sqrt{|\det(\sigma_Q)|}}\frac{\partial^{2N}}{\prod_{j=1}^M\partial^{n_j}\alpha_j\partial^{n_j}\alpha_j^*}e^{\frac{1}{2}\alpha^{T}K\alpha}\Bigg|_{\alpha_j=0}\\
        &=\frac{1}{\sqrt{|\det\sigma_Q)|}}\frac{\partial^{2N}}{\prod_{l=1}^N\partial\alpha_l\partial\alpha_l^*}e^{\frac{1}{2}\alpha^{T} K\alpha}\Bigg|_{\alpha_l=0},
    \end{split}
\end{align}
where in the second row the indices $l=1,\dots,N$ correspond to the output modes with $n_j=1$. To evaluate Eq.~(\ref{eq:expv5}), we introduce Fa\`a di Bruno's formula in Sec.~\ref{sec:bruno}, showing that it can be evaluated, as follows
\begin{align}\label{eq:as22}
    \begin{split}
        \text{Pr}(\bar{n})
        &=\frac{1}{\sqrt{|\det(\sigma_Q)|}}\sum_{j=1}^{(2N-1)!!}\prod_{k=1}^N K_{\mu_j(2k-1),\mu_j(2k)}\\
        &=\frac{1}{\sqrt{|\det(\sigma_Q)|}}\text{Haf}(K_S),
    \end{split}
\end{align}
where $\mu_j\in S_{2N}$ (symmetric group of $2N$ elements) define the indices of measured photons (bright modes) corresponding to perfect matching $j$ (of which there exist $(2N)!/(2^NN!)=(2N-1)!!$), and $K_S$ is the submatrix of $K$ corresponding to the indices of measured photons.

We observe that $\text{Pr}(\bar{n})$ is directly proportional to the number of perfect matchings associated with the observed modes with indices $\bar{n}$ when $A$ is the adjacency matrix of the complete graph of modes, $K=A^{\oplus 2}=c(A\oplus A)$, and $K_S$ the submatrix of $K$ corresponding to the observed modes. Therefore, the number of perfect matchings can be obtained by sampling from a Gaussian distribution with covariance $\sigma_Q=\sigma+I_{2M}/2$.

\section{Fa\`a di Bruno's formula}\label{sec:bruno}
Equation~(\ref{eq:expv4}) can be evaluated as discussed by Kruse et al~\cite{PhysRevA.100.032326}, using Fa\`a di Bruno's formula
\begin{align}\label{eq:dibruno}
    \frac{\partial^n}{\partial x_1\cdots\partial x_n}f(y)
    =\sum_{\pi\in P_n}f^{(\vert\pi\vert)}(y)\prod_{B\in\pi}\frac{\partial^{\vert B\vert}y}{\prod_{l\in B}\partial x_l},\quad
    y
    =y(x_1, x_2,\dots, x_n),
\end{align}
where ${P_n}$ is the set of partitions of $n$ indices $\{1,2,\dots,n\}$, while $\vert\pi\vert$ is the number of blocks of partition $\pi$, and $\vert B\vert$ is the number of elements in block $B$.

\paragraph{Example 1.} A simple example with $n=3$ illustrates Eq.~(\ref{eq:dibruno}) as applied to computing $\frac{\partial^3}{\partial x_1\partial x_2\partial x_3}f(y)$, with $y=y(x_1,x_2,x_3)$. The set $P_3$ of possible partitions of indices $\{1,2,3\}$ includes one partition $\pi_1$ with one block $\pi_1=\{\{1,2,3\}\}$ ($\vert\pi_1\vert=1$, one block of size three); three partitions with two blocks, including $\pi_2=\{\{1,2\},\{3\}\}$, $\pi_3=\{\{1,3\},\{2\}\}$, $\pi_4=\{\{2,3\},\{1\}\}$ ($\vert\pi_{2,3,4}\vert=2$); and finally one partition with three blocks $\pi_5=\{\{1\},\{2\},\{3\}\}$ ($\vert\pi_5\vert=3$). Hence, $|P_3|=5$ and we obtain
\begin{align}\label{eq:dibruno2}
    \begin{split}
        \frac{\partial^3}{\partial x_1\partial x_2\partial x_3}f(y) 
        &=f'(y)\frac{\partial^3 y}{\partial x_1\partial x_2\partial x_3}+\dots\\
        &f''(y)\left(\frac{\partial^2 y}{\partial x_1\partial x_2}\frac{\partial y}{\partial x_3}+\frac{\partial^2 y}{\partial x_1\partial x_3}\frac{\partial y}{\partial x_2}+\frac{\partial^2 y}{\partial x_2\partial x_3}\frac{\partial y}{\partial x_1}\right)+\dots\\
        &f'''(y)\frac{\partial y}{\partial x_1}\frac{\partial y}{\partial x_2}\frac{\partial y}{\partial x_3}.
    \end{split}
\end{align}
Analogously, to evaluate Eq.~(\ref{eq:expv5}), we use $n=2N$, $x=\alpha$, $y=\frac{1}{2}\alpha^{T}K\alpha$, and $f(y)=e^y$. Note that in this case $f^{(\vert\pi\vert)}(y)=f(y)$ for all partitions. It is important to note that the function $\frac12\alpha^{T}K\alpha$ is quadratic in $\alpha$, so all derivatives of third order or higher vanish. Furthermore, the argument of Eq.~(\ref{eq:expv5}) is evaluated at $\alpha_l=0$, so all first order derivatives also vanish. This leaves only the partitions for which $\vert B\vert=2$ for all blocks, which implies $\vert\pi\vert=N$ {(\em i.e.}, perfect matchings, the partitions of $2N$ indices into $N$ blocks of pairs can be interpreted as permutations of the $2N$ photon indices). Therefore, we obtain
\begin{align}\label{eq:ffdb1}
    \frac{\partial^n}{\partial x_1\cdots\partial x_n}f(y)
    =\frac{\partial^{2N}e^{\frac{1}{2}\alpha^{T}K\alpha}}{\prod_{l=1}^N\partial\alpha_l\partial\alpha_l^*}
    =e^{\frac{1}{2}\alpha^{T}K\alpha}\sum_{\pi\in PM}\prod_{B\in\pi}\frac{\partial^{2}y}{\prod_{l\in B}\partial x_l},
\end{align}
where $PM\subset P_{2N}$ denotes the subset of perfect matchings. Using that the second-order derivatives of $y$ can be expressed in terms of the matrix elements of $K$, we finally get
\begin{align}
    \begin{split}
        \text{Pr}(\bar{n})
        &=\frac{1}{\sqrt{|\det(\sigma_Q)|}}\frac{\partial^{2N}}{\prod_{l=1}^N\partial\alpha_l\partial\alpha_l^*}e^{\frac{1}{2}\alpha^{T} K\alpha}\Bigg|_{\alpha_l=0}\\
        &=\frac{1}{\sqrt{|\det(\sigma_Q)|}}\sum_{j=1}^{(2N-1)!!}\prod_{k=1}^N K_{\mu_j(2k-1),\mu_j(2k)}\\
        &=\frac{1}{\sqrt{|\det(\sigma_Q)|}}\text{Haf}(K_S),
    \end{split}
\end{align}
%

\begin{figure}[h]
    \begin{center}
        \includegraphics[width=0.64\textwidth]{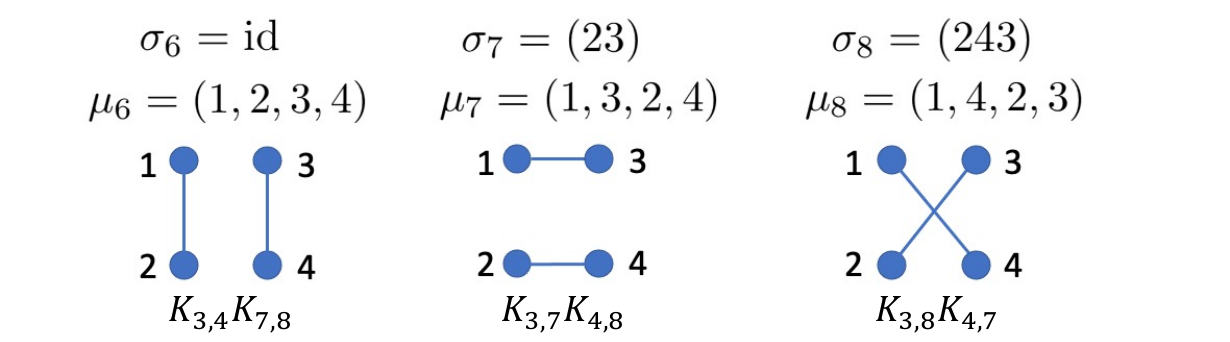}
        \includegraphics[width=0.64\textwidth]{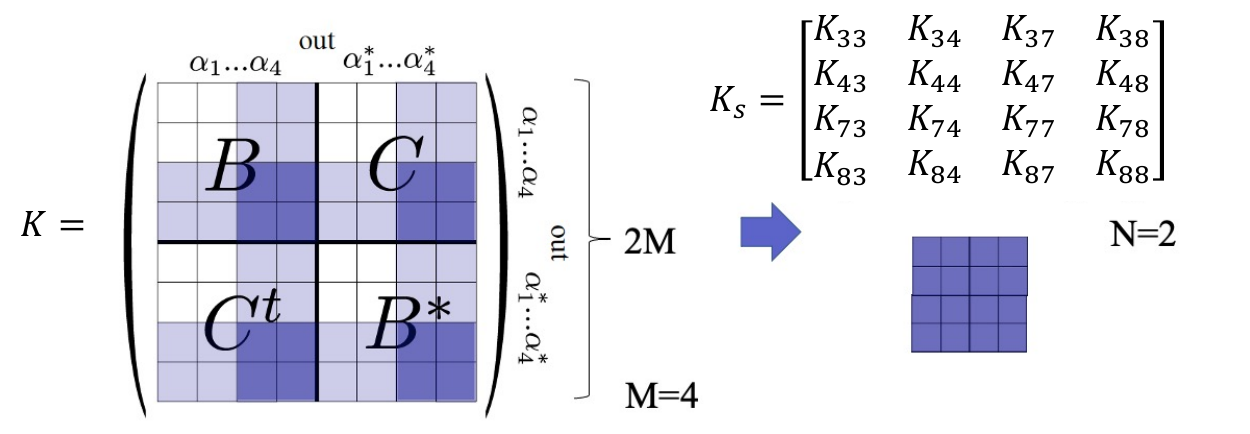}
    \end{center}
    \caption{Top: Index combinations as perfect matchings of a graph with four nodes. Bottom: Construction of the $2\times 2$ submatrix $K_S$ from the $8\times8$ symmetric matrix $K=A^{\oplus 2}$ for two photons measured in the last two output modes 3 and 4.}
    \label{fig:figbp}
\end{figure}

\paragraph{Example 2.}
A simple example, with $N=2$, shows how the partitions that contribute to Eq.~(\ref{eq:ffdb1}) can be interpreted as permutations of the photon indices.
Let us assume that $M=4$ and $N=2$ photons are measured in the last two output modes 3 and 4. We then have $\alpha=\{\alpha_3,\alpha_4,\alpha_3^*,\alpha_4^*\}$. We label the indices as follows: $\{\alpha_3\rightarrow 1,\alpha_4\rightarrow 2,\alpha_3^*\rightarrow 3,\alpha_4^*\rightarrow 4\}$. The perfect matchings are given by
\begin{itemize}
    \item $\pi_{PM_1}=\{\{1,2\},\{3,4\}\}$
    \item $\pi_{PM_2}=\{\{1,3\},\{2,4\}\}$
    \item $\pi_{PM_3}=\{\{1,4\},\{2,3\}\}$
\end{itemize}
We therefore conclude that
\begin{align}
    \begin{split}
        \sum_{\pi\in PM}\prod_{B\in\pi}\frac{\partial^{2}y}{\prod_{l\in B}\partial x_l}
        &=y^{(2)}_{1,2}y^{(2)}_{3,4}+y^{(2)}_{1,3}y^{(2)}_{2,4}+
        y^{(2)}_{1,4}y^{(2)}_{2,3}\\
        &=\frac{\partial^{2}y}{\partial\alpha_{3}\partial\alpha_{4}}\frac{\partial^{2}y}{\partial\alpha_{3}^{*}\partial\alpha_{4}^{*}}+\frac{\partial^{2}y}{\partial\alpha_{3}\partial\alpha_{3}^{*}}\frac{\partial^{2}y}{\partial\alpha_{4}\partial\alpha_{4}^{*}}+\frac{\partial^{2}y}{\partial\alpha_{3}\partial\alpha_{4}^{*}}\frac{\partial^{2}y}{\partial\alpha_{4}\partial\alpha_{3}^{*}}\\
        &=K_{3,4}K_{7,8}+K_{3,7}K_{4,8}+K_{3,8}K_{4,7}.   
    \end{split}
\end{align}
We note that the index combinations of the partial derivatives in the first row can be mapped into permutations corresponding to perfect matchings. These permutations (Fig.~\ref{fig:figbp}) can be written in vector format as $\mu_1=(1,2,3,4)$, $\mu_2=(1,3,2,4)$ and $\mu_3=(1,4,2,3)$, also denoted as $\sigma_1=\text{id}$, $\sigma_2=(23)$, and $\sigma_3=(243)$. Note, that the blocks are ordered from left to right corresponding to their block indices from lowest to highest and the numbers within a block are also in increasing order, so $\mu_j(2k-1)<\mu_j(2k+1)$ and $\mu_j(2k-1)<\mu_j(2k)$.
%

\section{Covariance matrix of squeezed and rotated state}\label{sec:sque-cov}
The goal of Appendices~\ref{sec:sque-cov} and~\ref{sec:K-N-rot} is to prove Eq.~\eqref{eq:kaa}, which establishes a relationship between the covariance matrix of the $M$-mode squeezed and rotated state, $\sigma$, and the graph adjacency matrix $A$. In this appendix, we first give the expression of $\sigma$ expressed according to the squeezing parameters and the $M$-mode rotation matrix $\mathbf{U}$ that defines the linear interferometer. Then in Appendix~\ref{sec:K-N-rot} we show how $A^{\oplus 2}=cA\oplus cA$ is identified as the so-called kernel matrix $K$, which one-to-one correspond to the covariance matrix $\sigma$.

Recall the definition of the squeezing operation
\begin{align}
    \hat{S}(r)
    =e^{(-r(\hat{a})^2+r(\hat{a}^{\dagger})^2)/2},
\end{align}
where the real-valued $r$ is referred to as the squeezing parameter. We aim to show that the action of the squeezing operator on the creation and annihilation operators are equivalent to the following linear Bogoliubov transformation
\begin{align}\label{eq:sque-bogo}
    \begin{bmatrix}
        \hat{a}\\
        \hat{a}^{\dagger}
    \end{bmatrix}
    \mapsto
    \begin{bmatrix}
        \text{cosh}(r) & \text{sinh}(r)\\
        \text{sinh}(r) & \text{cosh}(r)
    \end{bmatrix}
    \begin{bmatrix}
        \hat{a}\\
        \hat{a}^\dagger
    \end{bmatrix}
    =
    \begin{bmatrix}
        \hat{a}^{'}\\
        (\hat{a}')^{\dagger}
    \end{bmatrix}.
\end{align}
To proof this relationship, we derive the following
\begin{align}
    \hat{S}(r)^\dagger\hat{a}\hat{S}(r)
    =\text{cosh}(r)\hat{a}+\text{sinh}(r)\hat{a}^\dagger.
\end{align}
We set
\begin{align}
    \hat{A}=
    (r(\hat{a})^2-r(\hat{a}^{\dagger})^2)/2,
\end{align}
so that $\hat{S}(r)^\dagger=e^{\hat{A}}$. Then, using the Baker--Campbell--Hausdorff formula, we have
\begin{align}
    \begin{split}
        \hat{S}(r)^\dagger\hat{a}\hat{S}(r)
        &=e^{\hat{A}}\hat{a}e^{-\hat{A}}\\
        &=\sum_{k=0}^\infty\frac{1}{k!}[\hat{A},[\hat{A},\dots[\hat{A},\hat{a}]\dots]],
    \end{split}
\end{align}
where each term contains $k$ commutators. Noting that
\begin{align}
    \begin{split}
        [\hat{A},\hat{a}]
        &=\frac{1}{2}[r(\hat{a})^2-r(\hat{a}^{\dagger})^2,\hat{a}]
        =r\hat{a}^\dagger,\\
        [\hat{A},\hat{a}^\dagger]
        &=r\hat{a},
    \end{split}
\end{align}
we have
\begin{align}
    [\hat{A},\dots[\hat{A},\hat{a}]\dots]]
    =
    \begin{cases}
        r^k\hat{a} & \text{if $k$ is even},\\
        r^k\hat{a}^\dagger & \text{if $k$ is odd}.
    \end{cases}
\end{align}
Then we obtain
\begin{align}
    \begin{split}
        \hat{S}(r)^\dagger\hat{a}\hat{S}(r)
        &=\hat{a}\sum_{k=0}^\infty\frac{r^{2k}}{(2k)!}+\hat{a}^\dagger\sum_{k=0}^\infty\frac{r^{2k+1}}{(2k+1)!}\\
        &=\hat{a}\text{cosh}(r)+\hat{a}^\dagger\text{sinh}(r).
    \end{split}
\end{align}
We now show that the covariance matrix of single mode squeezed state is
\begin{align}\label{eq:squeez_cov}
    \mathbf{\sigma}'(r)
    =\frac{1}{2}
    \begin{bmatrix}
        \text{cosh}^2(r)+\text{sinh}^2(r) & 2\text{cosh}(r)\text{sinh}(r)\\
        2\text{cosh}(r)\text{sinh}(r) & \text{cosh}^2(r)+\text{sinh}^2(r)
    \end{bmatrix}.
\end{align}
For the first element of the squeezed state covariance matrix $\sigma_{11}'$, we substitute in the definition of the covariance matrix in Eq.~\eqref{eq:covar}
\begin{align}
    \begin{split}
        \sigma_{11}^{'}
        &=\frac{1}{2}\langle\{\hat{a}^{'},(\hat{a}')^{\dagger}\}\rangle\\
        &=\frac{1}{2}\langle\hat{a}^{'}(\hat{a}')^{\dagger}+(\hat{a}')^{\dagger}\hat{a}^{'}\rangle\\
        &=\frac{1}{2}\langle(\text{cosh}(r)^2+\text{sinh}(r)^2)(\hat{a}\hat{a}^\dagger+\hat{a}^\dagger\hat{a})+2\text{sinh}(r)\text{cosh}(r)(\hat{a}\hat{a}+\hat{a}^\dagger\hat{a}^\dagger)\rangle\\
        &=\frac{1}{2}(\text{cosh}(r)^2+\text{sinh}(r)^2)\langle\hat{a}\hat{a}^\dagger+\hat{a}^\dagger\hat{a}\rangle+\text{sinh}(r)\text{cosh}(r)\langle(\hat{a}\hat{a}+\hat{a}^\dagger\hat{a}^\dagger)\rangle\\
        &=\frac{1}{2}(\text{cosh}(r)^2+\text{sinh}(r)^2),
    \end{split}
\end{align}
where the last equality uses $\langle\hat{a}\hat{a}^\dagger+\hat{a}^\dagger\hat{a}\rangle=1$ and $\langle\hat{a}\hat{a}\rangle=\langle\hat{a}^\dagger\hat{a}^\dagger\rangle=0$ for the vacuum state. Note that all averages are taken for the vacuum state density matrix since the creation and annihilation operators are in the Heisenberg representation. Similarly,
\begin{align}
    \begin{split}
        \sigma_{12}^{'}
        &=\frac{1}{2}\langle 2\hat{a}^{'}\hat{a}^{'}\rangle\\
        &=\langle(\text{cosh}(r)\hat{a}+\text{sinh}(r)\hat{a}^\dagger)(\text{cosh}(r)\hat{a}+\text{sinh}(r)\hat{a}^\dagger)\rangle\\
        &=\langle\text{cosh}(r)^2(\hat{a}\hat{a})+\text{sinh}(r)^2(\hat{a}^\dagger\hat{a}^\dagger)+2\text{sinh}(r)\text{cosh}(r)\{\hat{a}^\dagger,\hat{a}\}\rangle\\
        &=\text{sinh}(r)\text{cosh}(r).
    \end{split}
\end{align}
The derivation for the other two elements of $\sigma^{'}$ are analogous. Gathering all elements we obtain Eq.~\eqref{eq:squeez_cov}.

Next, we show that applying an $M$-mode rotation, specified by an $M\times M$ unitary rotation matrix $U$, on the single-mode squeezed states specified by Eq.~\eqref{eq:squeez_cov}, results in a state with the following covariance matrix
\begin{align}
    \mathbf{\sigma}
    =
    \begin{bmatrix}
        U & 0\\
        0 & U^*
    \end{bmatrix}
    \sigma_{\text{sque}}
    \begin{bmatrix}
        U^* & 0\\
        0 & U
    \end{bmatrix}^T,
\end{align}
where $\sigma_{\text{sque}}$ is the generalization of Eq.~\eqref{eq:squeez_cov} to $M$ modes, as shown in Eq.~\eqref{eq:M_sque_cov}. To prove this, we first write a given $2M\times 2M$ covariance matrix $\tilde{\sigma}$ in the following block form
\begin{align}
    \tilde{\sigma}
    =
    \begin{bmatrix}
        B & G\\
        D & C
    \end{bmatrix},
\end{align}
where all four blocks are $M\times M$ matrices. We now show that an $M$-mode rotation specified by the $M\times M$ unitary matrix $U$ would rotate the covariance matrix to be
\begin{align}\label{eq:N_rot_cov}
    \tilde{\sigma}^{'}
    =
    \begin{bmatrix}
        U & 0\\
        0 & U^*
    \end{bmatrix}
    \tilde{\sigma}
    \begin{bmatrix}
        U^* & 0\\
        0 & U
    \end{bmatrix}^T
    =
    \begin{bmatrix}
        B^{'} & G^{'}\\
        D^{'} & C^{'}
    \end{bmatrix},
\end{align}
where $B^{'}=UBU^\dagger$, $C^{'}=U^*CU^T$, $G^{'}=UGU^T$, $D^{'}=U^*DU^\dagger$. As derived in Eq.~\eqref{eq:rot_aadag}, the $M$-mode rotation linearly combines all $M$ original annihilation/creation operators to obtain the rotated annihilation/creation operators, with the combination coefficients specified by elements of $U$
\begin{align}
    \begin{split}
        \hat{a}_k^{'}
        &=\sum_{j=1}^MU_{kj}\hat{a}_j,\\
        (\hat{a}_k')^{\dagger}
        &=\sum_{j=1}^M U_{kj}^*\hat{a}_j^\dagger.
    \end{split}
\end{align}
According to the definition of the covariance matrix in Eq.~\eqref{eq:covar}, an element $\sigma^{'}_{ij}$ of the upper-left block $B$ ($1\leq i,j\leq M$) of the rotated covariance matrix $\mathbf{\sigma}^{'}$, can be expanded as
\begin{align}
    \begin{split}
        \tilde{\sigma}_{ij}^{'}
        &=\left\langle\{\hat{a}_i^{'}(\hat{a}_j')^{\dagger}\}\right\rangle/2\\
        &=\left\langle\hat{a}_i^{'}(\hat{a}_j')^\dagger+(\hat{a}_j')^{\dagger}\hat{a}_i^{'}\right\rangle/2\\
        &=\frac12\left\langle\left(\sum_{l=1}^MU_{il}\hat{a}_l\right)\left(\sum_{m=1}^M U_{jm}^*\hat{a}_m^\dagger\right)+\left(\sum_{m=1}^M U_{jm}^*\hat{a}_m^\dagger\right)\left(\sum_{l=1}^M U_{il}\hat{a}_l\right)\right\rangle\\
        &=\frac12\sum_{l,m=1}^MU_{il}U_{jm}^*\left\langle\hat{a}_l\hat{a}_m^\dagger+\hat{a}_m^\dagger\hat{a}_l\right\rangle\\
        &=\sum_{l,m=1}^MU_{il}\tilde{\sigma}_{lm}U_{mj}^\dagger\\
        &=(UBU^\dagger)_{ij},
    \end{split}
\end{align}
which proves $B^{'}=UBU^\dagger$. The same procedure is carried out for $C$, $G$ and $D$ to prove Eq.~\eqref{eq:N_rot_cov}. In particular, with $\tilde{\sigma}=\mathbf{\sigma}_{\text{sque}}$, we have
\begin{align}\label{eq:n-rot-squeezed-cov_main2}
    \sigma_{\text{out}}
    =
    \begin{bmatrix}
        U & 0\\
        0 & U^*
    \end{bmatrix}
    \sigma_{\text{sque}}
    \begin{bmatrix}
        U^* & 0\\
        0 & U
    \end{bmatrix}^T,
\end{align}
as in Eq.~\eqref{eq:n-rot-squeezed-cov_main}.

\section{Kernel matrix for the N-mode rotated state}\label{sec:K-N-rot}
The goal of this section is to prove Eq.~\eqref{eq:kaa}. First, we show that the squeezed state covariance matrix $\mathbf{\sigma}'(r)$, derived as in Eq.~\eqref{eq:squeez_cov}, corresponds to the kernel matrix $K(r)$ that has the following form
\begin{align}
    K(r)
    =
    \begin{bmatrix}
        \text{tanh}(r) & 0\\
        0 & \text{tanh}(r)
    \end{bmatrix},
\end{align}
where the kernel matrix $K$ of a Gaussian state is defined according to its covariance matrix:
\begin{align}\label{eq:def_K}
    \begin{split}
        \mathbf{\sigma}_{Q}
        &=\mathbf{\sigma}+I/2,\\
        K
        &=X(I-\mathbf{\sigma}_{Q}^{-1}),\\
        X
        &
        =
        \begin{bmatrix}
            0 & 1\\
            1 & 0
        \end{bmatrix}.
    \end{split}
\end{align}
According to Eqs.~\eqref{eq:squeez_cov} and \eqref{eq:def_K},
\begin{align}
    \begin{split}
        \mathbf{\sigma}'_{Q}(r)
        &=\sigma'(r)+I/2\\
        &=\frac{1}{2}
        \begin{bmatrix}
            \text{cosh}^2(r)+\text{sinh}^2(r)+1 & 2\text{cosh}(r)\text{sinh}(r)\\
            2\text{cosh}(r)\text{sinh}(r) & \text{cosh}^2(r)+\text{sinh}^2(r)+1
        \end{bmatrix}\\
        &=
        \begin{bmatrix}
            \text{cosh}^2(r) & \text{cosh}(r)\text{sinh}(r)\\
            \text{cosh}(r)\text{sinh}(r) & \text{cosh}^2(r)
        \end{bmatrix}.
    \end{split}
\end{align}
Taking the inverse of the matrix,
\begin{align}
    \begin{split}
        (\mathbf{\sigma}_{Q}'(r))^{-1}
        &=\frac{1}{\text{cosh}^4(r)-\text{sinh}^2(r)\text{cosh}^2(r)}
        \begin{bmatrix}
            \text{cosh}^2(r) & -\text{cosh}(r)\text{sinh}(r)\\
            -\text{cosh}(r)\text{sinh}(r) & \text{cosh}^2(r)
        \end{bmatrix}\\
        &=\frac{1}{\text{cosh}^2(r)}
        \begin{bmatrix}
            \text{cosh}^2(r) & -\text{cosh}(r)\text{sinh}(r)\\
            -\text{cosh}(r)\text{sinh}(r) & \text{cosh}^2(r)
        \end{bmatrix}\\
        &=
        \begin{bmatrix}
            1 & -\text{tanh}(r)\\
            -\text{tanh}(r) & 1
        \end{bmatrix},
    \end{split}
\end{align}
and substitution in the definition for $K$ yields
\begin{align}
    \begin{split}
        K(r)
        &=
        \begin{bmatrix}
            0 & 1\\
            1 & 0
        \end{bmatrix}
        \left(
        \begin{bmatrix}
            1 & 0\\
            0 & 1
        \end{bmatrix}
        -
        \begin{bmatrix}
            1 & -\text{tanh}(r)\\
            -\text{tanh}(r) & 1
        \end{bmatrix}
        \right)\\
        &=
        \begin{bmatrix}
            \text{tanh}(r) & 0\\
            0 & \text{tanh}(r)
        \end{bmatrix}.
    \end{split}
\end{align}
The results above are for 1-mode squeezed states. Generalizing to $M$-single-mode squeezed states is trivial since all modes are unentangled with each other. In that case, we have
\begin{align}\label{eq:I-Nsq}
    I_{M}-\mathbf{\sigma}_{M-squeezed,Q}^{-1}
    =
    \begin{bmatrix}
        0 & \bigoplus_{j=1}^M\text{tanh}(r_j)\\
        \bigoplus_{j=1}^M\text{tanh}(r_j) & 0
    \end{bmatrix}\\,
\end{align}
and the $M$-mode squeezed state kernel matrix is
\begin{align}
    K_{M-squeezed}
    =
    \begin{bmatrix}
        \bigoplus_{j=1}^M\text{tanh}(r_j) & 0\\
        0 & \bigoplus_{j=1}^M\text{tanh}(r_j)
    \end{bmatrix}.
\end{align}
Now we show that the covariance matrix of the $N$-mode squeezed state in Eq.~\eqref{eq:n-rot-squeezed-cov_main2} corresponds to the following kernel matrix
\begin{align}
    K
    =c(A\oplus A)
    =:A^{\oplus 2},
\end{align}
where the adjacency matrix $A$ is decomposed according to Takagi's factorization as
\begin{align}
    A
    =U\left(\frac{1}{c}\bigoplus_{j=1}^M\text{tanh}(r_j)\right)U^T,
\end{align}
where $1/c$ is a constant greater than the largest singular value of $A$, ensuring that every scaled singular value can be represented by the form of $\text{tanh}(r_j)$. We begin by setting
\begin{align}
    \tilde{U}
    =
    \begin{bmatrix}
        U & 0\\
        0 & U^*
    \end{bmatrix},
\end{align}
so that Eq.~\eqref{eq:n-rot-squeezed-cov_main2} becomes
\begin{align}
    \mathbf{\sigma}
    =\tilde{U}\mathbf{\sigma}'(r)\tilde{U}^\dagger.
\end{align}
The corresponding matrix $\mathbf{\sigma}_{Q}=\mathbf{\sigma}+I/2$ can be written as
\begin{align}
    \begin{split}
        \mathbf{\sigma}_{Q}
        &=\mathbf{\sigma}+I/2\\
        &=\tilde{U}\mathbf{\sigma}'(r)\tilde{U}^\dagger +\tilde{U}I\tilde{U}^\dagger/2\\
        &=\tilde{U}\left(\mathbf{\sigma}'(r)+I/2\right)\tilde{U}^\dagger\\
        &=\tilde{U}\mathbf{\sigma}'_{Q}(r)\tilde{U}^\dagger.
    \end{split}
\end{align}
Moreover, since $\tilde{U}$ is unitary,
\begin{align}
    \mathbf{\sigma}_{Q}^{-1}
    =\tilde{U}\mathbf{\sigma}'_{Q}(r)^{-1}\tilde{U}^\dagger.
\end{align}
Therefore, 
\begin{align}
    \begin{split}
        I-\mathbf{\sigma}_{Q}^{-1}
        &=I-\tilde{U}\mathbf{\sigma}'_{Q}(r)^{-1}\tilde{U}^\dagger\\
        &=\tilde{U}\left(I-\sigma_Q'(r)^{-1}\right)\tilde{U}^\dagger\\
        &=\tilde{U}
        \begin{bmatrix}
            0 & \bigoplus_{j=1}^N\text{tanh}(r_j)\\
            \bigoplus_{j=1}^N\text{tanh}(r_j) & 0
        \end{bmatrix}
        \tilde{U}^\dagger\\
        &=
        \begin{bmatrix}
            U & 0\\
            0 & U^*
        \end{bmatrix}
        \begin{bmatrix}
            0 & \bigoplus_{j=1}^N\text{tanh}(r_j)\\
            \bigoplus_{j=1}^N\text{tanh}(r_j) & 0
        \end{bmatrix}
        \begin{bmatrix}
            U^\dagger & 0\\
            0 & U^T
        \end{bmatrix}\\
        &=
        \begin{bmatrix}
            0 & U\bigoplus_{j=1}^N\text{tanh}(r_j)U^T\\
            U^*\bigoplus_{j=1}^N\text{tanh}(r_j)U^\dagger & 0
        \end{bmatrix},
    \end{split}
\end{align}
where the third equality is obtained according to Eq.~\eqref{eq:I-Nsq}. The kernel matrix $K=X_{2M}\big(I_{2M}-\mathbf{\sigma}_{Q}^{-1}\big)$ is then
\begin{align}
    \begin{split}
        K
        &=
        \begin{bmatrix}
            0 & I_M\\
            I_M & 0
        \end{bmatrix}
        \begin{bmatrix}
            0 & U\bigoplus_{j=1}^M\text{tanh}(r_j)U^T\\
            U^*\bigoplus_j^M\text{tanh}(r_j)U^\dagger & 0
        \end{bmatrix}\\
        &=
        \begin{bmatrix}
            U^*\bigoplus_{j=1}^M\text{tanh}(r_j)U^\dagger & 0\\
            0 & U\bigoplus_{j=1}^M\text{tanh}(r_j)U^T
        \end{bmatrix}\\
        &=cA^*\oplus cA\\
        &=c(A\oplus A)\\
        &=A^{\oplus 2},
    \end{split}
\end{align}
since $A$ is real.

\bibliography{export.bib}
\end{document}